\documentclass[12pt]{article}
\usepackage{latexsym}
\usepackage{amssymb}
\usepackage{graphics}

\makeatletter
\@addtoreset{equation}{section}
\renewcommand{\theequation}{\thesection.\arabic{equation}}

\newcommand{\dd}{ \hbox{d} }
\def\bear{\begin{array}{l}}
\def\ear{\end{array}}

\newcommand{\nn}{\nonumber \\}

\font\zfont = cmss10 
\newcommand{\ZZ}{\hbox{\zfont Z\kern-.4emZ}}

\ifx\TwoupWrites\UnDeFiNeD\else\target{\magstepminus1}{11.3in}{8.27in}
\source{\magstep0}{7.5in}{11.69in}\fi
\newfont{\fourteencp}{cmcsc10 scaled\magstep2}
\newfont{\titlefont}{cmbx10 scaled\magstep3}
\newfont{\authorfont}{cmcsc10 scaled\magstep1}
\newfont{\fourteenmib}{cmmib10 scaled\magstep2}
\skewchar\fourteenmib='177
\newfont{\elevenmib}{cmmib10 scaled\magstephalf}
\skewchar\elevenmib='177
\makeatletter
\newcommand\nonsequentialeqnum{
\@addtoreset{equation}{section}
\def\theequation{\arabic{section}.\arabic{equation}}}
\newif\ifp@bblock \p@bblocktrue
\newcommand\nopubblock{\p@bblockfalse}
\newcommand\topspace{\hrule height 0pt depth 0pt \vskip}
\newcommand\p@bblock{\begingroup \tabskip=\hsize minus \hsize
\baselineskip=1.5\ht\strutbox \topspace-2\baselineskip
\halign to\hsize{\strut ##\hfil\tabskip=0pt\crcr
\the\Pubnum\crcr\the\date\crcr}\endgroup}

\renewcommand\titlepage{\ifx\TwoupWrites\UnDeFiNeD\null
\vspace{-1.7cm}\fi
\vskip0.6cm
\ifp@bblock\p@bblock \else\hrule height 0pt \relax \fi}
\makeatother
\newtoks\date
\newtoks\Pubnum
\newtoks\pubnum
\newcommand{\frontpageskip}{\vspace{12pt plus .5fil minus 2pt}}
\renewcommand{\title}[1]{\frontpageskip
\begin{center}{\titlefont #1}\end{center}\par}
\renewcommand{\author}[1]{\frontpageskip\par\begin{center}
{\authorfont #1}\end{center}
\nobreak
}

\renewcommand{\thanks}[1]{\footnote{#1}}
\renewcommand{\abstract}{\par\frontpageskip\centerline{
\fourteencp Abstract}
\vspace{8pt plus 3pt minus 3pt}}

\begin{document}

\begin{titlepage}
\hfill
\vbox{
    \halign{#\hfil         \cr
           hep-th/0612205  \cr
           } 
      }  
\vspace*{20mm}
\begin{center}
{\Large {\bf Relating Schwarzschild Black Holes to
Branes-Antibranes}\\} \vspace*{15mm}
{\sc Dan Gl\"uck}
\footnote{e-mail: {\tt gluckdan@post.tau.ac.il}}
and {\sc Yaron Oz}
\footnote{e-mail: {\tt yaronoz@post.tau.ac.il}}

\vspace*{1cm}
{\it {Raymond and Beverly Sackler Faculty of Exact Sciences\\
School of Physics and Astronomy\\
Tel-Aviv University , Ramat-Aviv 69978, Israel}}\\

\end{center}

\begin{abstract}
We construct in the supergravity framework a relation between thermal
chargeless non-extremal black three-branes and thermal Dirichlet
branes-antibranes systems. We propose this relation as a possible
explanation for the intriguing similarity between the black branes
Bekenstein-Hawking entropy and the field theory entropy of thermal
branes-antibranes. We comment on various relations between branes,
antibranes and non-BPS branes in type II string theories.

\end{abstract}
\vskip .7cm

December 2006

\end{titlepage}

\setcounter{footnote}{0}

\newpage

\section{Introduction and summary}

One of the challenges for any quantum theory of gravity is to
provide a microscopic description of the Bekenstein-Hawking entropy
of black holes. Such a description has been provided for certain
classes of extremal
\cite{Strominger:1996sh,Breckenridge:1996is,Maldacena:1996gb,Johnson:1996ga}
and near-extremal \cite{Callan:1996dv,Horowitz:1996fn,Gubser:1996de}
black holes in the framework of superstring theory. The black holes
in these examples are realized by D-branes, and the microscopic
degrees of freedom consist of the D-branes open strings. The
microscopic description of the Bekenstein-Hawking entropy of
non-extremal black holes remained largely as an open problem.

Brane systems and their open string degrees of freedom have been
shown  to approximately give the correct behavior of black branes
Bekenstein-Hawking entropy in certain cases \cite{Horowitz:1996ay,Danielsson:2001xe}.
However, it is not clear
why these systems of branes and their open string degrees of freedom
should be related to the black branes whose
entropies they are supposed to count.
The purpose of this paper is to make a step towards understanding the relation between branes-antibranes
systems and non-extremal black branes.

An intriguing similarity between the black brane Bekenstein-Hawking
entropy and the field theory entropy of thermal branes-antibranes
has been found in \cite{Danielsson:2001xe}. The
low-frequency absorption and emission probabilities are similar as well, and so are the entropies in the rotating and 
charged cases \cite{Danielsson:2001xe,Garcia:2004vx,Guijosa:2004zn}. 
Similar works have been done for $p \ne 3$ \cite{Saremi:2004pi,Bergman:2004tz} and multicharged black holes
\cite{KalyanaRama:2004fk}.
An overview of these models as well as a detailed discussion 
appears in \cite{Garcia:2004vx}.

In the following we
will review the arguments of  \cite{Danielsson:2001xe}.
Consider a thermal system of $N$
$D3-\bar{D}3$ pairs of branes with temperature $T$, and $g_s N \gg 1$.
The mass squared of the strings stretched between the branes and the antibranes,
and in particular the tachyon $t$,
is expected to receive at weak coupling a correction of order 
$\Delta m^2 \sim g_s N T^2$. At strong coupling we parameterize the correction by
\begin{equation}
\Delta m^2 \sim (g_s N)^\alpha T^2
\end{equation}
with $\alpha$ some positive number. 
The mass squared of the tachyon is then 
\begin{equation}
m_t^2 \sim (g_s N)^\alpha T^2 - l_s^{-2} \ .
\end{equation}
Thus the tachyon mass becomes
much larger than $T$ in the regime 
\begin{equation}
\# (g_s N)^{-\alpha/2} l_s ^{-1} < T \ll l_s^{-1} \label{condition}
\end{equation}
where $\#$ stands for an unknown numerical constant.
The first inequality implies that the open strings stretched between D-branes 
and the anti D-branes are 
massive and the system is stable. Effectively, the open strings
attached to the D-branes are decoupled from the open strings
attached to the anti D-branes, and the open strings degrees of
freedom can be counted by summing up those attached to the D-branes
and those attached to the anti D-branes \cite{Danielsson:2001xe}.
Moreover, the second inequality implies that we can ignore the
massive open string modes and count only the massless ones, thus
counting the field theory degrees of freedom. 
We will take $g_s \ll 1$ in order to work at string tree level,
which means that $N\gg 1$.

Since the system consists of two decoupled subsystems, one of
D-branes and the other of anti D-branes, one can calculate its
energy $M$ and entropy $S$ as functions of $N$, $T$ and the
3-worldvolume $V$. Pairs of D-brane and anti D-brane can still be
created and annihilated. However, they are not emitted as closed
strings but rather as open strings since $g_s\ll 1$. The number of
pairs $N$ is related to the temperature $T$, and maximizing  the
entropy $S(M,N,V)$ with respect to $N$, we obtain the entropy in the
microcanonical ensemble $S(M,V)$. It turns out that this result is,
up to a  factor of $2^{3/4}$, the same as the entropy of a chargeless black
3-brane as a function of the mass and 3-volume
\cite{Danielsson:2001xe}. The temperature of the $D3-\bar{D3}$
branes system $T= (\partial M/\partial S)_V$ is $(g_s N)^{-1/4} l_s^{-1}$ 
up to a numerical factor. Thus, one finds that a thermal
$D3-\bar{D}3$ branes system in the regime $g_s N \gg 1$ and at
temperature $T=(g_s N)^{-1/4} l_s ^{-1}$ (up to a numerical factor)
is at thermal equilibrium, and has the same entropy $S(M,V)$ as a
black 3-brane, up to a  factor of $2^{3/4}$.

Since the system turns out to be stable only at temperature 
$T \sim (g_s N)^{-1/4} l_s ^{-1}$, 
the analysis is self-consistent only if this 
temperature turns out to be in the regime (\ref{condition}). This is 
satisfied only if $\alpha \ge 1/2$. If $\alpha = 1/2$, the consistency check 
requires a calculation of the exact numerical coefficient in (\ref{condition}).
In the weakly coupled  regime ($g_s N \ll 1$) $\alpha=1$,
but for $g_s N \gg 1$ the value of $\alpha$ is unknown.

\begin{figure}
\centerline{\includegraphics{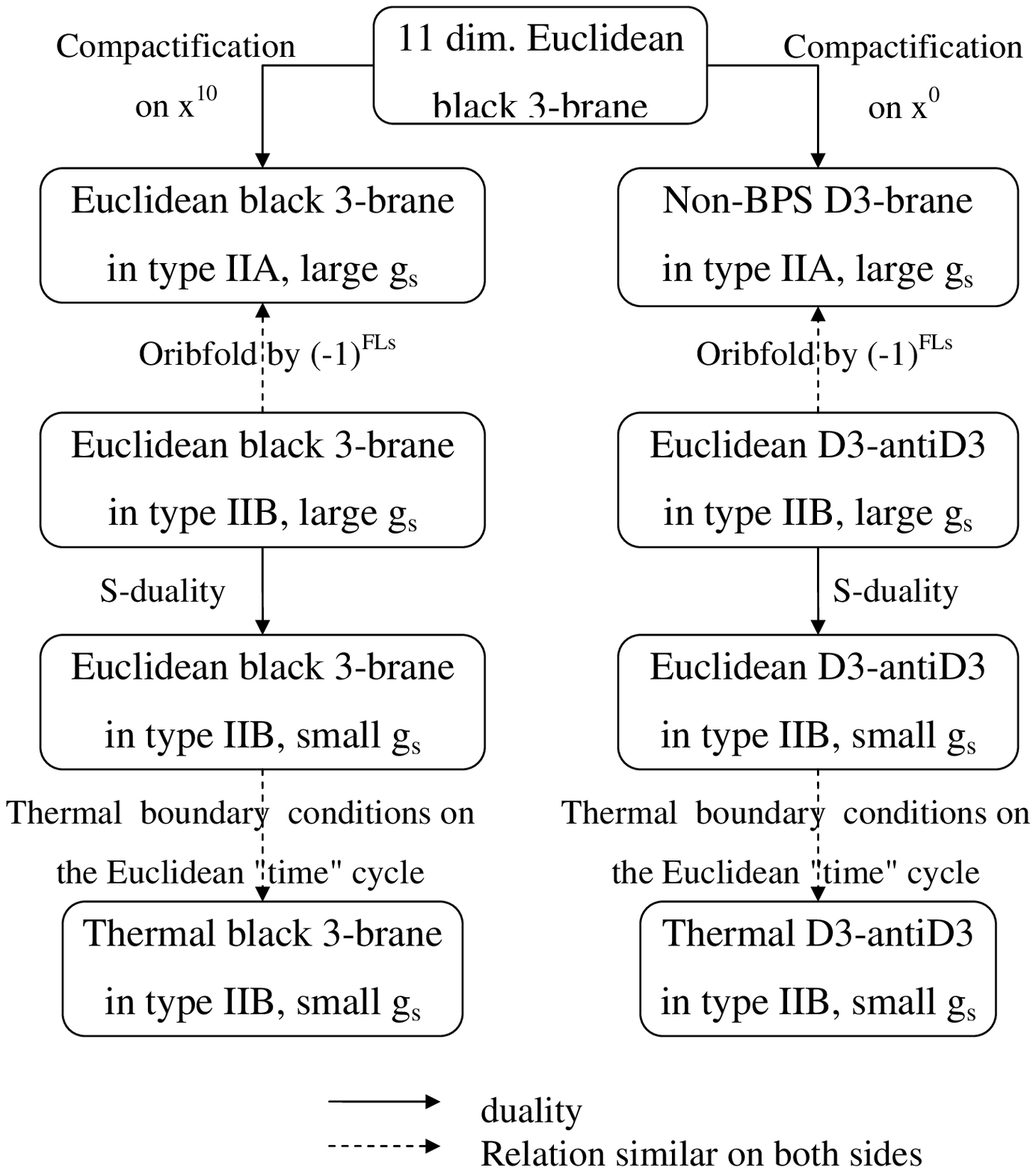}}
\caption{Relating thermal black 3-branes and the $D3-\bar{D3}$ branes system}
\label{fig1}
\end{figure}

The aim of this paper is to propose a possible explanation for this
intriguing similarity between the black branes Bekenstein-Hawking
entropy and the field theory entropy of thermal branes-antibranes.
We will work in the supergravity framework and construct a relation
between thermal chargeless non-extremal black three-branes and
thermal Dirichlet branes-antibranes systems. The relation that we
find is depicted in figure (\ref{fig1}).

The paper is organized as follows.\\
In section 2 we analyze supergravity backgrounds candidate for describing Dp branes-antibranes and
Dp non-BPS branes.
This is done by considering the space of supergravity solutions with the appropriate symmetries,
constructed in \cite{Zhou:1999nm} and studied in \cite{Brax:2000cf}.
All these solutions posses a naked curvature singularity. When considering
type IIA supergravity we can
distinguish "bad" naked curvature singularities from "good" ones \cite{Gubser:2000nd}
by lifting to eleven dimensions. We find
that for each $p$ there is precisely one solution with the appropriate symmetries and no "bad" naked 
singularity. This solution is the dimensional reduction of a bubble solution in eleven dimensions.
It is a limit of a family of solutions which describe D-branes wrapped on 
a non-supersymmetric cycle
\cite{Sen:1997pr,Lu:2006rb}, and we identify it with the Dp branes-antibranes (or non-BPS branes)
system.

In section 3 we discuss descent relations among
type IIA and type IIB D-brane systems \cite{Sen:1998ex,Sen:1999mg,Thompson:2001rw}.
These relations arise from orbifoldings by $(-1)^{{F_s}_L}$, where ${F_s}_L$ is the
spacetime fermion number arising from the left worldsheet sector. This orbifolding maps the type IIA superstring
theory to type IIB and vice versa. Its effect on $N$ brane-antibrane pairs or $N$
non-BPS branes has been discussed for $N=1$ in \cite{Sen:1998ex,Sen:1999mg}. We review this discussion
and generalize the study to $N>1$. We find that there are different ways to
realize such an orbifold, one of which does not change the charge of the solution. We discuss
how such an orbifold should be defined in superstring theory on a background of a chargeless supergravity
solution. We argue that the orbifold action can be defined so that the
form of the solution (i.e. the metric and dilaton) remains unchanged, perhaps up to changing the ADM mass
by a numerical factor.

In section 4 we discuss in detail the relation between the supergravity solution describing
non-extremal black 3-branes and the supergravity solution, which we proposed as a description of thermal
D3 branes-antibranes system. This relation is depicted in figure (\ref{fig1}).
In particular we assume that descent relations between type IIA and type IIB 
brane systems hold at large $g_s$.
We also find that both the black 3-branes and the branes-antibranes system
annihilate to closed strings at related limits. We remark on the degrees of freedom of the black 3-brane and 
about charged black 3-branes, still far from extremality.
In section 5 we discuss possible generalizations of this relation.

In Appendices A, B and C we provide details of certain aspects of time-independent supergravity
solutions with an $ISO(p)\times SO(9-p)$ symmetry.
In appendix D we comment on systems with $N$ non-BPS branes for $N>1$.
In appendices E and F we study certain suggestions regarding the field theory of the D3 branes-antibranes systems.

\section{The supergravity description of $Dp-\bar{Dp}$ and non-BPS $Dp$-branes}

In this section we consider time-independent chargeless asymptotically flat
type II supergravity solutions with an
ISO(p)$\times$SO(9-p) symmetry. We show that there are exactly two classes of solutions
that do not have
"bad" naked curvature singularity \cite{Gubser:2000nd}. One class is the
Horowitz-Strominger black $p$-branes. We identify the second class as the
$Dp-\bar{Dp}$ or non-BPS $Dp$-branes solution, depending on whether $p$ is even or odd and whether we
consider type IIA or type IIB supergravity.
In type IIA solutions of the latter class are dimensional reduction of eleven dimension "bubble" solutions 
\cite{}.

The $Dp-\bar{Dp}$ (non-BPS $Dp$) branes are solutions of type IIB (IIA) superstrings for odd
$p$, and vice versa for even $p$. They have an ISO(1,p)$\times$SO(9-p) symmetry and
are chargeless. In order to identify the corresponding type II supergravity solutions, we consider all asymptotically flat
time-independent chargeless solutions with this symmetry.
In fact we generalize the discussion and consider all
chargeless asymptotically flat
time-independent solutions with ISO(p)$\times$SO(9-p) symmetry. These have been constructed
in \cite{Zhou:1999nm} and include four parameters:
$r_0$, $c_1$, $c_2$ and $c_3$. One combination of the parameters is related to the ADM mass and
$c_3$ is related to the charge.
The chargeless solution in the Einstein frame reads
\begin{equation}
\bear
\dd s^2 =  ({f_- \over f_+})^{ (7-p) (3-p) \, c_1 + (2(7-p) +{1\over 4} (3-p)^2) c_2
-  4 (7-p) c_3 k \over  32 }  \left(- ({f_- \over f_+})^{-c_2}\dd t^2 +
(\dd x^i)^2 \right) \\
+ ({f_- \, f_+})^{ 2\over 7-p  }
({f_- \over f_+})^{- (p+1)(3-p)\, c_1   +  {1\over 4} (3-p)^2 c_2
+ 4 (p+1) c_3 k \over 32 } \left( \dd r^2 + r^2 \dd \Omega_{8-p}^2 \right)\\
e^{\phi-\phi_{\infty}} = ({f_- \over f_+})^ { {7-p \over 64} \left(4 (p+1) \, c_1  - (3-p) c_2 \right)
+ { 3-p \over 4} c_3 k} \\
f_\pm \equiv 1 \pm \left( r_0 \over r \right)^{7-p}\\
k  \equiv   \sqrt{- c_1^2 + { 1\over 4}\,  \left( {3-p \over 2} \, c_1 +
{ 7-p  \over 8}c_2 \right)^2 + 2{8-p\over 7-p} - { 7\, c_2^2\over 16}}  \, \, , \\
\ear
\label{Poincaremetric}
\end{equation}
i=1,...p, and $c_3 = \pm 1$.
The solution is invariant under a simultaneous flip of the signs of all four parameters
$c_1$, $c_2$, $c_3$ and ${r_0}^{7-p}$. We will therefore assume in the following that ${r_0}^{7-p}>0$.

The black $p$-branes solutions have
\footnote{The isotropic coordinates cover twice the region outside the horizon of the black brane,
and do not cover the region behind the horizon. Indeed the solution is invariant under
$r\rightarrow {r_0}^2/r$, so that the $r>r_0$ and $r<r_0$ regions cover the same region in spacetime.
It is geodetically incomplete at $r=r_0$. In the region behind the horizon $dr$ is timelike and isotropic
coordinates cannot be defined.}
 \cite{Zhou:1999nm,Horowitz:1991cd}
\begin{equation}
(c_1,c_2,c_3)=({3-p \over 2 (7-p)},-2,-1) \ .
\label{regsol}
\end{equation}
All other solutions for $1<p<7$, and in particular, all the chargeless solutions with 
ISO(1,p)$\times$SO(9-p) symmetry (i.e. with $c_2=0$)
have a naked curvature singularity at $r=r_0$.

When  $p=-1$, $c_2$ must be set to zero and (\ref{regsol}) cannot be
satisfied. Thus there is no chargeless black $-1$-brane.
$c_1$ is redundant and so there is only one solution in (\ref{Poincaremetric}), which we will 
mention later. When $p=0$ the parameter $c_2$ is redundant and 
(\ref{regsol}) is equivalent to $(c_1,c_2,c_3)= ({12\over 7}+{3\over 4}c_2,c_2,-1)$. As for 
$1<p<7$, this is the only solution with no naked curvature singularity. 
When $p=1$ there is one other solution, in addition to (\ref{regsol}), with no 
naked singularity, where $(c_1,c_2,c_3)=(7/6,2,-1)$. This is the nine-dimensional Euclidean
Schwarzschild black hole with a flat time coordinate.

A solution with naked curvature singularity is often pathological since the
singularity can be seen by asymptotic observers and its resolution may affect the whole spacetime.
Obviously, since the curvature corrections near the singularity are large, the supergravity approximation
is not valid.
Criteria for distinguishing pathological ("bad") from non-pathological ("good") naked singularities in
asymptotically $AdS_5$ backgrounds have been proposed in \cite{Gubser:2000nd}.
There, it has been
suggested that solutions with naked curvature singularities are pathological if they are
neither a limit of a family of solutions with no naked curvature singularity nor a dimensional
reduction of such a solution. It has also been argued that the nature of
naked curvature singularities may be independent of the spacetime asymptotics.

In some cases, naked curvature singularities can be resolved by incision of the geometry in a particular
radius where some cycle shrinks to a string scale size, while the curvature remains small
\cite{Johnson:1999qt,Johnson:2001wm}. For the
supergravity approximation to remain valid asymptotically, and in particular for winding states to be negligible
asymptotically, that cycle must be asymptotically large in string units. Therefore, for such a
mechanism to work, the supergravity solution must have two different length scales (in addition to
the asymptotic cycle scale), one determines the curvature at the incision radius and the
other determines the cycle scale there \footnote{In \cite{Johnson:1999qt,Johnson:2001wm}
these length scales
where $r_6$ and $r_2$, or $r_5$ and $r_1$.}.
Note that if the asymptotic volume of the $T^4$ in
\cite{Johnson:1999qt,Johnson:2001wm} is arbitrarily large in string  units, then the geometry is
arbitrarily close to the BPS $D6$ (or $D5$) geometry, which satisfies the criteria of
\cite{Gubser:2000nd} for a "good" singularity.
However, in the solutions (\ref{Poincaremetric})
there is only one such length scale, $r_0$, so the metric cannot be resolved
by such an incision.

Type IIA supergravity
solutions can be thought of as a dimensional reduction from eleven dimensions, and the eleven-dimensional
solution may have no naked curvature singularity. This happens precisely for just one solution for
every $-1 \le p < 7$, that is for
\begin{equation}
(c_1,c_2,c_3)=({12 - 2 p \over 7-p},0,sign(2-p)) \ .
\label{goodsing}
\end{equation}
Note that for $p=2$
the two signs of $c_3$ are equivalent, for $p=0$  the parameter $c_2$ is redundant and
the solution is equivalent to $(c_1,c_2,c_3)= ({12\over 7}+{3\over 4}c_2,c_2,1)$. For
$p=-1$, $c_2$ must be set to zero and $c_1$ is redundant. Thus, there is only one solution
up to changing the sign of $c_3$, which is equivalent to changing the sign of $r_0$.
This solution is
a dimensional reduction of the eleven-dimensional black hole.
We will argue in the next subsection that these type IIA solutions (\ref{goodsing}) describe
the $Dp-\bar{Dp}$ (non-BPS Dp-branes) for even (odd) $p$.

Finally, note that for the solutions with
\begin{equation}
(7-p)(3-p) c_1 + (2(7-p) +{1\over 4} (3-p)^2 -32) c_2 -  4 (7-p) c_3 k >0 \ ,
\end{equation}
the naked
singularity is infinitely redshifted with respect to an asymptotic observer.
Thus, we can ask whether such a singularity is "bad" as
it seems that regions far away from the singularity may not be significantly influenced by it.
However, the particle production rate of Bekenstein-Hawking radiation
measured by an asymptotic observer, calculated in appendix B, is infinite. Alternatively,
by analytically continuing the metric to Euclidean signature, there is no way to define a temperature
so that the singularity is resolved. This can be understood by noting that all the solutions which have
a naked curvature
singularity in the string frame, which means large $\alpha^\prime R$,
have a naked curvature singularity in the Einstein frame as well, which dictates the temperature.
Nevertheless, upon lifting to eleven dimensions, temperature can sometimes be defined 
so that spacetime is smooth, as will be
demonstrated below.
We may interpret this as a strong coupling effect.
As is shown in appendix B, among the solutions with an $ISO(p+1)\times SO(9-p)$
symmetry, this can be done
only for two sets of solutions, which turn out to have a flat time direction
in eleven dimensions.
One set of solutions is (\ref{goodsing}) and the other, given
by (\ref{goodsing}) with changing the signs of $c_1$ and $c_3$, is
having negative ADM mass and therefore unphysical.

\subsection{Interpreting the "good" supergravity solutions}

In the following we will argue that the type IIA solutions (\ref{goodsing}) describe
the $Dp-\bar{Dp}$ (non-BPS Dp-branes) for even (odd) $p$.
At first sight, one may hope to distinguish the branes-antibranes supergravity
solutions from the rest by the  energy-momentum tensor. Computing the components of the
energy-momentum tensor at the singularity in $r=r_0$ can be done by using the method of
\cite{Das:1997tk}. The branes-antibranes system at zero temperature is expected
to obey the relation $T_{ij}= - T_{00}\delta_{ij}$ \cite{Danielsson:2001xe}. However we show in
appendix A, that this relation holds for all the chargeless supergravity solutions with the appropriate
symmetry $ISO(p+1)\times SO(9-p)$.

The type IIA solutions (\ref{goodsing}) have been considered
in \cite{Lu:2006rb}. Their lift to eleven dimensions is the Euclidean
black p-branes background with a flat time direction. These solutions are known as the (static) bubble solutions.

For $p=6$, the ten-dimensional solution is the $D6-\bar{D6}$ system \cite{Sen:1997pr},
which is the $a\rightarrow 0$ limit of a family of solutions of branes and the antibranes separated
by a distance of $2a$ (for large $a$). The branes-antibranes
worldvolume  is $S^0\times T^{6}\times R$ for any $a\ne 0$.
For $p<6$, the ten-dimensional solutions have been shown to be a limit of a family
of solutions of D6-branes with an $S^{6-p}\times T^p \times R$ worldvolume topology  \cite{Lu:2006rb}.
Therefore, these solutions describe D-branes configurations on which open strings can end.
We identify them with Dp branes-antibranes (non-BPS Dp-brane) for even (odd) $p$, which are the
objects in type IIA superstring theory that are chargeless and posses the same symmetry  ISO(1,p)$\times$SO(9-p)
and on which open strings can end.
Similar interpretation has been proposed in \cite{Lu:2006rb}.
Another possible interpretation of this configuration is D6-branes wrapping a non-supersymmetric 
vanishing cycle.

We may argue why Dp branes-antibranes system for even $p$ can be expected to arise as a limit
of a family of solutions with an $S^{6-p}\times T^p \times R$ topology.
Dp BPS branes (non-BPS Dp-branes) for even (odd) $p<6$ can be thought of as
solitonic solutions of a $D6-\bar{D6}$ adjacent pair \cite{Sen:1999mg,Witten:1998cd}. Dp BPS
branes source a non-zero $C_{p+1}$-form, which arises in describing them as
solitons from the term
\begin{equation}
\int C_{p+1}\wedge F ... \wedge F
\end{equation}
in the $D6$-brane worldvolume action, and from a similar term, with an
opposite sign, in the $D6$-antibrane worldvolume action.

Thus, a Dp-BPS brane is an ${\cal M}_{6-p}\times T^p \times R$ submanifold of the $D6-\bar{D6}$,
with
\begin{equation}
\int_ {{\cal M}_{6-p}} F_1 \wedge ... F_1 - \int_ {{\cal M}_{6-p}} F_2 \wedge ... F_2 \ne 0
\end{equation}
where $F_1$ is the gauge field strength of the D6-brane
and $F_2$ the gauge field strength of the D6-antibrane.
The Dp brane-antibrane pair corresponds to two such solitons with opposite charges, in the limit where they
are on top of each other, and is having the same topology.

One may hope to be able
to identify open strings stretched between the branes as M2-branes from the M-theory point of view, as
in \cite{Sen:1997pr,Sen:1997kz}. However open strings which were identified so were only those
stretched between different branes which were located in different points in space. The geometries
we have considered for $p<6$ did not include different branes located in different points in space,
and therefore it may be impossible to make a similar identification here.

The bubble solution is unstable to small perturbations
\cite{Harmark:2005pp}. Thus the supergravity solutions which correspond 
to the $Dp-\bar{Dp}$ system is unstable at large $g_s$, which is hardly surprising.
This generalizes the instability of the $D6-\bar{D6}$ solution to small perturbations
\cite{Sen:1997pr,Dowker:1995gb,Dowker:1995sg}. In the following we ignore these instabilities, just as
we ignore the black three-brane instability to small perturbations \cite{Gregory:1994bj}. It
would be interesting to consider what is the effect of these instabilities on backgrounds on a thermal time cycle,
but this is beyond the scope of this work.

\subsection{Comparing to the boundary states approach}
We now compare our results to those obtained by a different approach, which uses branes boundary
states \cite{Brax:2000cf,Bertolini:2000jy} (for an interpretation of black p-branes
based on this approach see \cite{Bai:2006vv}).

The BPS brane boundary state receives contributions from two closed string sectors,
NSNS and RR. The non-BPS brane boundary state receives contribution only from the NSNS sector,
which is identical to the NSNS contribution of the BPS Dp-brane,
up to a factor which reflects the different tensions of the two \cite{Bergman:1998xv}.

For the BPS brane, the boundary state couplings to a graviton, a dilaton and an RR form have been
shown to be equal (up to an overall factor) to the sub-leading terms\footnote{This is the first correction
to a flat background, which is $\sim 1/r^{7-p}$ or, in Fourier transform, $\sim V_p /{k_\perp}^2$,
where $k_\perp$ is the momentum transverse to the brane.}
of the asymptotic metric, dilaton and RR form, respectively. This reflects the fact that due to
closed-open string duality, the coupling of a closed string to the brane can also be calculated from
supergravity for on-shell states \cite{DiVecchia:1997pr}.

Now consider the non-BPS brane. It has the same contribution from the NSNS sector as the
BPS brane (up to a factor), so we may expect that the sub-leading term of the asymptotic metric and
dilaton would be the same as in the BPS brane, up to an overall factor. Thus a stack of non-BPS branes
of ADM-mass $M$ may be expected to have the same asymptotic metric and dilaton (including
the sub-leading term) as a stack of BPS branes of the same ADM mass. This means that
$(c_1,c_2,c_3)=(0,0,-1)$ in (\ref{Poincaremetric}) for every $p$, as has been shown by an
equivalent method (using a BPS Dp-brane probe) in \cite{Brax:2000cf} (see also \cite{Bertolini:2000jy}).
This is a different point in the parameter space than (\ref{goodsing}), the two coinciding only for $p=6$, 
in which case they coincide with \cite{Sen:1997pr} as well.

However, for the chargeless supergravity solutions in question, there is no limit in which
the supergravity solution is decoupled from the asymptotic regime, and thus no limit in which open and
closed strings decouple \footnote{In
later sections we use the tools of \cite{Danielsson:2001xe} to discuss such an approximate duality in a
finite temperature case, for a particular temperature.}. We show this qualitatively in Appendix C (in
terms of the effective potential felt by a minimally coupled scalar).
In particular, for the non-BPS brane boundary state, the couplings to the closed string fields are
expected to receive higher loop corrections, which vanish in the BPS case.
Thus for large $g_s N$, where supergravity may be valid, the result can be different from the
naive boundary state calculation. Indeed we have
seen that the $(c_1,c_2,c_3)=(0,0,-1)$ supergravity solution (for $p\ne 6$) has a "bad" naked
singularity.

\section{Descent relations among supergravity solutions}

In this section we discuss the effect of taking a $(-1)^{{F_s}_L}$ orbifold of chargeless
supergravity solutions.
We argue that in these solutions the background fields do not change as a result
of this orbifolding (except for a possible change in the ADM mass). We use this in the next section to
propose that the black 3-brane in type IIB,
orbifolded by $(-1)^{{F_s}_L}$, gives the black 3-brane in type IIA. 
Orbifold in terms of open strings, as well as non-BPS Dp-branes, are defined at $g_s=0$,
and throughout this section we take $g_s$ to be zero or small.
However in other sections we assume that there is a generalization of these to large $g_s$ as well, and 
in particular that the $(-1)^{{F_s}_L}$ symmetry is non anomalous.
We will begin with a brief review
of known results.

\subsection{Orbifolding by $(-1)^{{F_s}_L}$: a brief review}

Type IIA and type IIB string theories are related by orbifolding by  $(-1)^{{F_s}_L}$, where
${{F_s}_L}$ is the spacetime fermion number coming from fields which are left-moving on the
worldsheet \cite{Sen:1998ex}. Let us first summarize some of the results described in
\cite{Sen:1998ex,Sen:1999mg,Thompson:2001rw}.

The action of $(-1)^{{F_s}_L}$ flips signs of $p$-forms and therefore takes a D-brane to
an anti D-brane and vice versa. A $Dp-\bar{Dp}$ pair is invariant under this action.
The open string spectrum of a $Dp-\bar{Dp}$ pair has 4 sectors, two of which (corresponding to
open strings with both ends attached to the brane or both to the antibrane) have the GSO projection
$(-1)^F = 1$, where $F$ is the worldsheet fermion number,
and the other two (corresponding to open strings with one end attached to the brane
and the other to the antibrane) have the GSO projection $(-1)^F = -1$. In particular the lowest mass
content consists of two real gauge fields degrees of freedom and two real tachyonic degrees of freedom
(i.e. a single complex tachyon). The four sectors can be described by a $2\times 2$ hermitian 
CP (Chan-Paton) matrix.

Orbifolding a type IIA (IIB) theory with a $Dp-\bar{Dp}$ pair by $(-1)^{{F_s}_L}$ changes
the theory to type IIB (IIA) with a non-BPS Dp-brane. Thus the effect of the orbifold on the open string 
spectrum can be described by a matrix $S$ acting on the CP matrix $\Lambda$ as: 
$\Lambda \rightarrow S \Lambda S^{-1}$, where
$S$ is either $\sigma^1$ or (equivalently) $\sigma^2$ and the remaining degrees of freedom after the
orbifolding have CP matrices $1$ and $S$. Let us take $S=\sigma^1$.

In the type IIB (IIA) obtained in this way, RR closed strings are in the twisted sector. Therefore
an RR insertion in the worldsheet creates a branch cut. If the worldsheet has a boundary, then the
boundary on the two sides of the branch cut would be attached, in the language of the original type IIA
(IIB), to a brane on one side and to an antibrane on the other side. The point where the branch
cut meets the boundary must have a
$\sigma^1$ CP factor. This means that an RR $p$-form insertion may have a non-vanishing 
(off-shell) two point function with an open string with a $\sigma^1$ CP factor, which indeed it 
has \cite{Sen:1999mg}.

By further orbifolding a type IIB (IIA) theory on a non-BPS Dp-brane by $(-1)^{{F_s}_L}$, one gets
a type IIA (IIB) theory on a BPS $Dp$-brane (or an antibrane).
This is because the type IIB (IIA) RR closed string is odd under the orbifold action, which
implies that an open string with a $\sigma^1$ CP factor is odd as well.
The only open string sector which survives the orbifolding is the one with CP matrix 
$1$, which has the GSO projection $(-1)^F = 1$, and so the open string spectrum is identical to that of
a BPS $Dp$-brane (or an antibrane).

To summarize, taking the $(-1)^{{F_s}_L}$ orbifold twice on a
$Dp-\bar{Dp}$ pair results in a BPS $Dp$-brane (or an antibrane), while the closed string theory
(either type IIA or type IIB) returns to itself.

\subsection{$(-1)^{{F_s}_L}$ orbifolds of $N$ brane-antibrane pairs}

For $N$ $Dp- \bar{Dp}$ pairs with $N>1$, taking the $(-1)^{{F_s}_L}$
orbifold once, we get $N$ non-BPS $Dp$-branes. We will now show that taking the
$(-1)^{{F_s}_L}$ orbifold again takes us to a system of $N_1$ $Dp$-branes and
$N_2\equiv N-N_1$ $Dp$-antibranes, where the value of $N_1$ depends on the realization
of the orbifold action.
In appendix D we propose a possible interpretation in terms of non-BPS brane orientation.

Consider a system of $N$ $Dp$-branes and $N$ $Dp$-antibranes in type IIA ($p$ is even).
The following argument can be repeated with type IIA and type IIB interchanged (for odd $p$).
The open string spectrum of a system of $N$ $Dp$-branes and $N$ $Dp$-antibranes on top of each
other can be written as a $2N\times 2N$ hermitian matrix
\begin{equation}
\left(
\begin{array}{cc}
A & T \\
\bar{T}& B
\end{array}
\right) \,  ,
\end{equation}
where $A$ ($B$) are in the adjoint representation of the $U(N)$ gauge group on the branes
(antibranes) and have GSO projection $(-1)^F = 1$,
while $T$ and $\bar{T}$ describe the open strings attached between the branes and the
antibranes, transforming in the bi-fundamental representation of $U(N)\times U(N)$ and
have GSO projection $(-1)^F = -1$, thus including $N^2$ complex tachyonic degrees of freedom.

The action of $(-1)^{{F_s}_L}$ is implemented in the open string spectrum by a
matrix $S_1$, taking any CP matrix $\Lambda$ to $S_1 \Lambda {S_1}^{-1} $.
$S_1$ inverts branes and antibranes and we may take it to be
\begin{equation}
S_1 = \left(
\begin{array}{cc}
0 & 1_{N\times N} \\
1_{N\times N}& 0
\end{array}
\right) \, .
\end{equation}

Only states even under $(-1)^{{F_s}_L}$ survive orbifolding by it, and these are
states with $A=B$ and $T=\bar{T}$. Thus we are left with states whose CP matrix is of the form
\begin{equation}
\left(
\begin{array}{cc}
A & T \\
T & A
\end{array}
\right)
\label{ATTA}
\end{equation}
with $A$ and $T$ hermitian. Note that now there are $N^2$ real tachyonic degrees of freedom,
and the open string spectrum is that of $N$ non-BPS $Dp$-branes of type IIB.

A type IIB RR-field (such as the $p$-form) is in the twisted sector. Thus its vertex operator
creates a branch cut in the worldsheet. For a worldsheet with a boundary, at the point where the branch
cut meets the boundary there should be a CP matrix:
\begin{equation}
C \equiv \left(
\begin{array}{cc}
0 & 1_{N\times N} \\
1_{N\times N} & 0
\end{array}
\right)
\label{CPform}
\end{equation}
This is because the $(-1)^{{F_s}_L}$ action inverts type IIA branes and antibranes.
This means that the RR $p$-form may have a non-vanishing (off-shell) coupling to a tachyon on
the non-BPS Dp-branes with the CP factor (\ref{CPform}). This coupling is indeed non-zero, since
D$(p-1)$ BPS branes can be described as solitonic solutions to the tachyon potential on the non-BPS
D$p$-branes, and the $p$-form should therefore be coupled to the derivative of one of the tachyonic
degrees of freedom which live on the non-BPS Dp-brane \cite{Sen:1999mg}.

Let us now consider the action of a second $(-1)^{{F_s}_L}$, orbifolding by which takes the type IIB
back to type IIA.
This $(-1)^{{F_s}_L}$ is implemented in the open string spectrum by a matrix $S_2$, taking any CP
matrix $\Lambda$ to $S_2 \Lambda {S_2}^{-1}$.
This $(-1)^{{F_s}_L}$ has the following properties: its square is the identity,
Type IIB RR fields are odd under it and therefore so are states with CP factor (\ref{CPform}), and 
it does not mix $A$ and $T$ in (\ref{ATTA}).
Therefore $S_2$ is
equivalent\footnote{Equivalence between two choices of $S_2$ means that the same
$(-1)^{{F_s}_L}$ action is induced by both.} to a matrix of the following form
\begin{equation}
\left(
\begin{array}{cc}
X & 0 \\
0 & -X
\end{array}
\right)
\end{equation}
where $X$ is an $N\times N$ matrix with eigenvalues $\pm 1$.
We may diagonalize $X$ and get the diagonal
$(+1,+1...,+1,-1...,-1)$, with $N_1$ times the eigenvalue $+1$ and $N_2\equiv N-N_1$ times the
eigenvalue $-1$, for some $0\le N_1 \le N$.

The open string degrees of freedom which are left invariant by this $(-1)^{{F_s}_L}$ consist of
${N_1}^2+{N_2}^2$ real degrees of freedom from the sector whose GSO projection is
$(-1)^F = 1$,
and $2 N_1 N_2$ real degrees of freedom from the sector whose GSO projection is $(-1)^F = -1$.
This
is the open string spectrum of a system of $N_1$ branes and $N_2$ antibranes (or $N_1$ antibranes
and $N_2$ branes: there is
a two-fold ambiguity, resembling the symmetry $N_1\leftrightarrow N_2$ which is a consequence of
the equivalence of $S_2$ and $-S_2$).

\subsection{Orbifolding by $(-1)^{{F_s}_L}$  in the supergravity framework}

For large $N$ (in fact, for every even $N$)
the orbifolding procedure can be defined so that the result of subsequent
orbifoldings on a chargeless system (i.e. one with equal number of branes and antibranes) is again a
chargeless system.
In the supergravity limit $g_s N \gg 1$, for a solution with no RR background fields,
such an orbifolding procedure does not create RR background fields. We will henceforth use 
this prescription for the $(-1)^{{F_s}_L}$ orbifold.
Orbifolding twice takes us from one $Dp-\bar{Dp}$ solution to another, so both should correspond to
the same supergravity solution (i.e. the same metric and dilaton), except for a change in the
ADM mass: The total number of brane-antibrane pairs is cut by two after every two subsequent
orbifoldings, and this seem to mean that the mass of the system is cut by half as well. In the $N=1$
case this is manifested by the fact that the tension of the brane is multiplied by a $1/\sqrt{2}$ factor 
after performing the orbifold once \cite{Sen:1999mg}. 
This analysis assumes $g_s\sim 0$, so we have $N\sim \infty$, and it may be
speculated that the correct way to take the supergravity limit is to have $N$ unchanged by the orbifold procedure.
If this is true, then orbifolding twice does not change the background fields at all.
Such a result may be desired from the closed string point of view, at least for small asymptotic string coupling 
$g_s$, as we will now explain.

If we take a $Z_2$ orbifold of a $Z_2$ orbifold of any string theory, where the second $Z_2$
is defined by flipping the sign of all twisted states of the first $Z_2$, then
the torus partition
function is the same as that of the original theory \cite{Ginsparg:1988ui}.
For a string theory whose worldsheet field theory is not
free, such as string theory on a curved background, the second $Z_2$ orbifold should be defined 
via the boundary conditions on the cycles of the worldsheet torus in
the torus partition function, rather than via its action on the spectrum. This is because the torus
partition function can no longer be computed as a sum over oscillators.

For a $(-1)^{{F_s}_L}$ orbifold, the twisted states are the RR and R-NS sectors, and the theory
has a new $(-1)^{{F_s}_L}$ symmetry which flips the sign of these. Thus, taking
the $(-1)^{{F_s}_L}$ orbifold twice on a closed string theory is precisely the
$Z_2$ orbifold of a $Z_2$ orbifold we have just described, and this gives us the original
theory back, as is well known for the flat background cases \cite{Sen:1999mg,Thompson:2001rw}.

Let us consider backgrounds of the type (\ref{Poincaremetric}).
A worldsheet action formulation of string theory on these curved backgrounds is unknown,
but far away from the singularity there may be such a formulation which does not involve open strings, but 
still accounts for the sub-leading (i.e. $\sim 1/r^{7-p}$) term of the asymptotic background fields. 
Then closed string scattering amplitudes,
computed to this sub-leading order of the background geometry, are expected to remain unchanged after
orbifolding the theory twice. Therefore these sub-leading terms of the background fields should
remain unchanged as well. For the solutions
(\ref{Poincaremetric}) the whole geometry can be deduced from these sub-leading terms,
and so the whole geometry should not change after taking the $(-1)^{{F_s}_L}$ orbifold
twice.
It is then natural to assume that one $(-1)^{{F_s}_L}$ orbifold does not change
the metric as well, although it takes us from type IIA to type IIB and vice versa (as noted in section 2,
both supergravity theories have the same set of chargeless solutions).

In the rest of this paper we assume, for simplicity, that the background including the ADM mass remains unchanged
under the $(-1)^{{F_s}_L}$ orbifolding procedure. However the assumption that the ADM mass does
not change is not crucial to our following arguments, since our results hold only up to a numerical factor,
so they remain similar even if the ADM mass does change by a numerical factor as a result of the 
orbifolding procedure.

\section{ $D3-\bar{D3}$ and black 3-branes}

Let us start with the following (non-supersymmetric) supergravity solution in 11 dimensions
with time coordinate Wick rotated (i.e. a Euclidean solution)
\begin{eqnarray}
\dd s^2 &=& f(r) \dd {x_0}^2 +
f^{-1}(r) \dd r^2
+ r^2 \dd {\Omega_5}^2
+\,\sum_{i=1}^3 \dd {x_i}^2
+ \dd {x_{10}}^2
\nn
A_{\lambda\mu\nu}&=& 0 \nn
f(r) &\equiv& 1-\left({r_0\over r}\right)^4
\label{11d}
\end{eqnarray}

We will assume that $r_0 \gg {M_{11}}^{-1}$ ($M_{11}$ being the eleventh dimensional 
Planck mass).
$r$ takes the values $r\ge r_0$ and $x_0$ is periodic with a period $2\pi R_0$ with $R_0 = r_0 /2$,
so that the metric has no conical singularity at $r=r_0$.
We will also assume $x_{10}$ and the $x_i$-s are periodic with periods $2\pi R_{10}$
and $2\pi R_{i}$ respectively, where
$R_{1,2,3,10} \gg {M_{11}}^{-1}$.
Under these conditions, eleven dimensional supergravity is a good 
approximation for M-theory, on a Euclidean background.

In the following section, we will reduce the solution either along the $x_0$ or the $x_{10}$ cycle to
obtain two dual Euclidean type IIA supergravity solutions. Each reduced solution is an orbifold
of a different type IIB solution by $(-1)^{{F_s}_L}$, where we will pick the realization
for this orbifold as described in the previous section. We will get two related type IIB
Euclidean solutions.
By performing S-duality we will arrive at a
weakly coupled black 3-brane solution of type IIB supergravity
on one side, and a Euclidean version of what we interpret as the $D3-\bar{D3}$ system on the other
side.
Finally we will change the identification along the $x_{10}$ or $x_0$ cycle (the one we did not reduce
along) as to include the action of $(-1)^{F_s}$,
where $F_s$ is the spacetime fermion number. Thus we
switch from modding the theory by translations along the full
$x_0$ or $x_{10}$ cycle, which we will denote $P_0$ or $P_{10}$, respectively,
to modding it by $P_0\cdot (-1)^{F_s}$ or $P_{10}\cdot (-1)^{F_s}$, respectively.
This changes the Euclidean solution to a thermal one, and we end up with a
thermal black 3-brane on one side, and a thermal $D3-\bar{D3}$ system
on the other side. The two are not dual, since the last orbifolding
is done along cycles which are different in the original 11 dimensional theory.
The two solutions have no fermionic background fields, so the background fields do not change.
Let us see what implications this may have on the relation between the entropies of the
two systems.

The Bekenstein-Hawking entropy depends only on the
background fields, and this suggests that moving from the Euclidean supergravity
solutions to the thermal ones have no effect on the entropy comparison we are making.
Thus the black 3-brane and the $D3-\bar{D3}$ should have the same entropy.
For the $D3-\bar{D3}$ we are actually calculating the entropy by using a field theory,
and for completeness we should also examine the effect of
changing the boundary conditions from Euclidean to thermal on the field theory entropy computation.

The effective field theory that describes the thermal $D3-\bar{D3}$, and that is used for calculating its
entropy, is strongly coupled ($g_s N \gg 1$). This theory (which is basically the same as the one on
$D3$s) admits a relation between
strong coupling and weak coupling, and in particular the entropies are
related \cite{Gubser:1996de}. In the weakly coupled theory (weakly coupled $N=4$ SYM) the
entropy comes from the degrees of freedom of the gauge field supermultiplet. The bosonic degrees
of freedom are not affected by changing the boundary conditions from Euclidean to thermal (i.e.
adding a $(-1)^F$ on the Euclidean time cycle). Since these contribute half of the entropy,
such a change in the boundary conditions will change the calculated entropy by a factor of two at most.
Due to the relation between the entropies in weak and the strong couplings, this suggests that
the strong coupling entropy also changes by a numerical factor at most when moving from
Euclidean to thermal boundary conditions.
%
%

Thus we may expect the black 3-brane Bekenstein-Hawking entropy and the $D3-\bar{D3}$ entropy
to be of the same order of magnitude.
%

\subsection{Reduction along $x_{10}$: thermal black 3-branes}

Reducing (\ref{11d})
along the $x_{10}$ cycle
gives the following type IIA
supergravity solution (in Einstein frame)
\begin{eqnarray}
\dd s^2 &=& (M_{11}R_{10})^{1/4}
\left(f(r) \dd {x_0}^2 +
f^{-1}(r) \dd r^2
+ r^2 \dd {\Omega_5}^2
+\,\sum_{i=1}^3 \dd {x_i}^2 \right)
\nn
e^{\phi}&=& (R_{10}M_{11})^{3/2} \gg 1
\nn
l_s &=& {R_{10}}^{-1/2}{M_{11}}^{-3/2}
\end{eqnarray}
with no $p$-forms.
This is the Euclidean black 3-brane solution of type IIA.

The $R_{10}$ dependence of the metric and dilaton can be understood by noting that we may
redefine $x_{10}\rightarrow (M_{11}R_{10})^{-1}x_{10}$ before the reduction, so that
the periodicity would be independent of $R_{10}$; This would give $g_{10,10}=
(M_{11}R_{10})^2$.

Note that the string coupling is large. However we know that this background will get no
large corrections because it is a dimensional reduction of the eleven-dimensional background,
where supergravity is a good approximation (thus the large string coupling may
give large corrections to the spectrum of the theory but not to the background fields).

This solution can be thought of as
the result of
orbifolding a type IIB string theory, with the same background fields, by
$(-1)^{{F_s}_L}$. Indeed, we have argued in the last section that the background is invariant under
this operation, and the string theory changes from type IIB to type IIA. The type IIB solution is a
Euclidean black 3-brane.
This is S-dual to the black 3-brane solution with the same metric
in Einstein frame and string length, but with $g_s = e^{\phi} = (R_{10}M_{11})^{-3/2} \ll 1$.

The metric in string frame is now
\begin{eqnarray}
\dd s^2 &=& (M_{11}R_{10})^{-1/2}
\left(f(r) \dd {x_0}^2 +
f^{-1}(r) \dd r^2
+ r^2 \dd {\Omega_5}^2
+\,\sum_{i=1}^3 \dd {x_i}^2 \right)
\end{eqnarray}

By the following redefinitions
\begin{eqnarray}
r \rightarrow (M_{11}R_{10})^{-1/4}r &,&
r_0 \rightarrow r_{BH} \equiv (M_{11}R_{10})^{-1/4}r_0
\nn
x_0 \rightarrow \tau \equiv (M_{11}R_{10})^{-1/4}x_0 &,&
x_i \rightarrow (M_{11}R_{10})^{-1/4}x_i
\end{eqnarray}
for $i=1,2,3$, one arrives at the usual Euclidean
black 3-brane solution, with no awkward factors
\begin{eqnarray}
\dd s^2 &=&
\widetilde{f}(r) \dd \tau ^2 +
\widetilde{f}^{-1}(r) \dd r^2
+ r^2 \dd {\Omega_5}^2
+\,\sum_{i=1}^3 \dd {x_i}^2
\nn
\widetilde{f}(r) &\equiv& 1-\left({r_{BH}\over r}\right)^4
\label{Schwarzschild}
\end{eqnarray}

with $\tau$ and the new $x_i$ having periods
$\tau_{BH}\equiv 2\pi (M_{11}R_{10})^{-1/4}R_0$ and $2\pi (M_{11}R_{10})^{-1/4}R_i$,
respectively.

Its Schwarzschild radius is large in string length units:
$r_{BH}/l_s = r_0 {R_{10}}^{1/4}{M_{11}}^{5/4}\gg 1$.
Therefore the curvature is small everywhere and supergravity is a good approximation.

Switching from this theory (which has a $P_0$ symmetry, where $P_0$ is the translation
$\tau \rightarrow \tau+\tau_{BH}$) to
a theory orbifolded by $P\cdot (-1)^{F_s}$, with the same background fields,
we end up with a thermal black 3-brane in type IIB
with temperature, coupling and string length
\begin{eqnarray}
T_{BH} &=& 1/\tau_{BH}= (M_{11}R_{10})^{1/4}/(2\pi R_0)
\nn
{g_s}_{BH} &=& e^{\phi} = (R_{10}M_{11})^{-3/2} \ll 1
\nn
{l_s}_{BH} &=& {R_{10}}^{-1/2}{M_{11}}^{-3/2}
\label{TglSch}
\end{eqnarray}
Its Schwarzschild radius is $r_{BH}=(M_{11}R_{10})^{-1/4}r_0$, so the temperature
satisfies the usual relation (for a black 3-brane)
$T_{BH}=1/(\pi r_{BH})$.

The 3-volume and the ADM-mass of the black brane are\footnote{We are using here
the conventions of \cite{Brax:2000cf}.}
\begin{eqnarray}
{V_3}_{BH} &=&  (2\pi)^3 (M_{11}R_{10})^{-3/4} \prod_{i=1}^3 R_i
\nn
M_{BH} &=&  \alpha_0 V_3 {r_{BH}}^4/ {g_s}^2 {l_s}^8
\nn
&=& 16 \alpha_0 (2\pi)^3 (M_{11}R_{10})^{-7/4} {R_0}^4
{R_{10}}^{7}{M_{11}}^{15}\prod_{i=1}^3 R_i
\nn
\alpha_0 &\equiv& {5 \omega_5 \over (2\pi)^7 } = {5 \over 2^7 \pi^4 }
\label{MassSch}
\end{eqnarray}
with $\omega_5=\pi^3$ the volume of a 5-sphere of unit radius.

The temperature, mass, 3-volume and Schwarzschild radius have been given in length units which are
normalized so that the string frame metric is asymptotically $\eta_{\mu\nu}$, which is the correct
normalization to be used when comparing supergravity and field theory,
as in \cite{Itzhaki:1998dd,Maldacena:1997cg} \footnote{For comparison, length units which are
normalized so that the Einstein frame metric is asymptotically $\eta_{\mu\nu}$, are related to these
by a factor of ${g_s}^{-1/4}$.}.

\subsection{Reduction along $x_{0}$: $D3-\bar{D3}$}

Reducing (\ref{11d})
along the $x_{0}$ cycle
gives the following type IIA
supergravity solution (in Einstein frame)
\begin{eqnarray}
\dd s^2 &=& (M_{11}R_0)^{1/4}
\left[f^{1/8} (r) \left( \dd {x_{10}}^2
+ r^2 \dd {\Omega_5}^2
+\,\sum_{i=1}^3 \dd {x_i}^2 \right)
+ f^{-7/8}(r) \dd r^2
\right]
\nn
e^{\phi}&=& (R_{0}M_{11})^{3/2}f^{3/4}
\nn
l_s &=& {R_{0}}^{-1/2}{M_{11}}^{-3/2}
\end{eqnarray}
with no $p$-forms.

The $R_0$ dependence can again be understood by noting that we may
redefine $x_0\rightarrow (M_{11}R_0)^{-1}x_0$ before the reduction, so that
the periodicity would be independent of $R_0$, and that would give $g_{00}= (M_{11}R_0)^2 f(r)$.

The Lorentzian form of this solution (with $\tau\rightarrow i t$)
is the solution (\ref{Poincaremetric}) with (\ref{goodsing}) for $p=3$ (as can be seen after a coordinate
transformation as in \cite{Lu:2006rb}).
We have argued in section 2 that this describes
the non-BPS D3-brane of type IIA. As has been mentioned earlier, it has been shown in
\cite{Sen:1998ex}
that the non-BPS D3-brane of type IIA is the result of orbifolding the type IIB $D3-\bar{D3}$ system
by $(-1)^{{F_s}_L}$.
The latter is S-dual to a $D3-\bar{D3}$ system with the same $l_s$ and Einstein frame metric, but with
$e^{\phi}= (R_{0}M_{11})^{-3/2}f^{-3/4}$.

Its metric in string frame is 
\begin{eqnarray}
\dd s^2 &=& (M_{11}R_0)^{-1/2}
\left[f^{-1/4} (r) \left( \dd {x_{10}}^2
+ r^2 \dd {\Omega_5}^2
+\,\sum_{i=1}^3 \dd {x_i}^2 \right)
+ f^{-5/4}(r) \dd r^2
\right]
\end{eqnarray}

By the following redefinitions
\begin{eqnarray}
r \rightarrow (M_{11}R_0)^{-1/4}r &,&
r_0 \rightarrow r_{D\bar{D}} \equiv (M_{11}R_0)^{-1/4}r_0
\nn
x_{10} \rightarrow \tau \equiv (M_{11}R_0)^{-1/4}x_{10} &,&
x_i \rightarrow (M_{11}R_0)^{-1/4}x_i
\end{eqnarray}
for $i=1,2,3$, one arrives at the following solution
\begin{eqnarray}
\dd s^2 &=& \hat{f}^{-1/4} (r) \left( \dd \tau^2
+ r^2 \dd {\Omega_5}^2
+\,\sum_{i=1}^3 \dd {x_i}^2 \right)
+ \hat{f}^{-5/4}(r) \dd r^2
\nn
e^{\phi}&=& (R_0 M_{11})^{-3/2}\hat{f}^{-3/4}
\nn
\hat{f}(r) &\equiv& 1-\left({r_{D\bar{D}}\over r}\right)^4
\label{DDbar}
\end{eqnarray}
with $\tau$ and $x_i$ having periods
$\tau_{D\bar{D}}\equiv 2\pi (M_{11}R_0)^{-1/4}R_{10}$ and $2\pi (M_{11}R_{0})^{-1/4}R_i$,
respectively.

Switching from this theory (which has a $P_{10}$ symmetry, where $P_{10}$ is the translation
$\tau \rightarrow \tau+\tau_{D\bar{D}}$) to
a theory orbifolded by $P_{10}\cdot (-1)^{F_s}$, with the same background fields,
we end up with a thermal $D3-\bar{D3}$ system
with temperature, asymptotic coupling and string length
\begin{eqnarray}
T_{D\bar{D}} &=& 1/\tau_{D\bar{D}} = (M_{11}R_0)^{1/4}/(2\pi R_{10})
\nn
{g_s}_{D\bar{D}} &\equiv& e^{\phi_{\infty}} = (R_0 M_{11})^{-3/2} \ll 1
\nn
{l_s}_{D\bar{D}}  &=& {R_0}^{-1/2}{M_{11}}^{-3/2}
\label{TglDDbar}
\end{eqnarray}

The 3-volume and the ADM mass of the system are
\begin{eqnarray}
{V_3}_{D\bar{D}} &=&  (2\pi)^3 (M_{11}R_0)^{-3/4} \prod_{i=1}^3 R_i
\nn
M_{D\bar{D}} &=&  \alpha_0 V_3 {r_{D\bar{D}}}^4/ 8 {g_s}^2 {l_s}^8
\nn
&=& 2 \alpha_0 (2\pi)^3 (M_{11}R_{0})^{-7/4}
{R_0}^{11}{M_{11}}^{15}\prod_{i=1}^3 R_i
\nn
\label{MassDDbar}
\end{eqnarray}
with $\alpha_0$ defined in (\ref{MassSch}); The $1/8$ factor in the ADM mass comes from the $1/8$
power of $\hat{f}(r)$ in $g_{tt}$ in the Einstein frame.
Again, length units are normalized so that the string frame metric is asymptotically $\eta_{\mu\nu}$.

\subsection{Comparison of entropies}

The entropy of a black 3-brane with a 3-volume $V_{BH}$
and an ADM-mass $M_{BH}$ is \cite{Danielsson:2001xe,Peet:1997es}\footnote{Note that
$(\partial S/\partial M)_V = \pi r_{BH} = 1/T_{BH}$.}
\begin{equation}
S_{BH} = 2^{15\over 4} \, 5^{-{5\over 4}} \, \pi^2
\, {{g_s}_{BH}}^{1/2} \, {{l_s}_{BH}}^2
\, {V_{BH}}^{-{1\over 4}} \, {M_{BH}}^{5\over 4}
\label{SBH}
\end{equation}

The entropy of a thermal $D3-\bar{D3}$ system with a 3-volume $V_{D\bar{D}}$
and an ADM-mass $M_{D\bar{D}}$ was argued in \cite{Danielsson:2001xe} to be (assuming
(\ref{condition}))
\begin{equation}
S_{D\bar{D}} = (n_b/6)^{1\over 4} \, 2^3 \, 5^{-{5\over 4}} \, \pi^2
\, {{g_s}_{D\bar{D}}}^{1/2} \, {{l_s}_{D\bar{D}}}^2
\, {V_{D\bar{D}}}^{-{1\over 4}} \, {M_{D\bar{D}}}^{5\over 4}
\label{SDDbar}
\end{equation}
where $n_b$ is the number of bosonic degrees of freedom in the (strongly coupled)
field theory living on each of the decoupled $D3$s and $\bar{D3}$s, and is assumed to be $n_b=6$ by
comparison with strongly coupled field theory on $D3$s alone.
This estimate was done by using a field theory description which was justified in
\cite{Danielsson:2001xe}. The energy and temperature of the field theory are assumed to coincide with
the mass and temperature of the $D3-\bar{D3}$ supergravity solution.
Note however that there is no exact
open string - closed string duality, and in principle the energies and temperatures of both sides may differ.

The thermal equilibrium temperature
of the $D3-\bar{D3}$ system is \cite{Danielsson:2001xe}
\begin{equation}
T_{equ} = (\partial M_{D\bar{D}}/ \partial S_{D\bar{D}})_{V_{D\bar{D}}}
= 2^{3/2} / \pi r_{D\bar{D}}
\label{Tequ}
\end{equation}
It is easy to verify that $r_{D\bar{D}} \gg {l_s}_{D\bar{D}}$.

It is easy to see that if we neglect numerical factors, then
by setting $R_{10}=R_0$,
the mass, 3-volume, string length and $g_s$, as well as the temperature,
will be equal in both systems.
Additionally, the temperature is of the same order as (\ref{Tequ}), so that
the condition (\ref{condition}) is satisfied and the $D3-\bar{D3}$ system is close to equilibrium.
Since(\ref{SBH}) and (\ref{SDDbar}) are identical
as functions of $M$, $V$, $g_s$ and $l_s$, we get the same entropy
for both systems (neglecting numerical factors), as expected.

Note that by relating $R_0$ and $R_{10}$, the three parameters of the black hole
$r_{BH}$, ${g_s}_{BH}$ and ${l_s}_{BH}$ are no longer
independent, so we consider only a two-parameter subspace of the full three-parameter space.
However we are trying to match up with Bekenstein-Hawking's formula, which
depends only on a two-parameter space as well (since $g_s$ and $l_s$ do not appear in the formula
separately but only in the combination $g_s {l_s}^4$).

We now turn to a precise comparison of the entropies of both systems.
substituting for the values found in (\ref{TglSch},\ref{MassSch}, \ref{TglDDbar},
\ref{MassDDbar}) we get for (\ref{SBH}) and (\ref{SDDbar})
\begin{eqnarray}
S_{BH} &=& 2^3 {R_0}^5 {R_{10}}^5 {M_{11}}^{13} \prod_{i=1}^3 {R_i}
\nn
S_{D\bar{D}} &=& 2^{-{3\over 2}} {R_0}^{10} {M_{11}}^{13} \prod_{i=1}^3 R_i
\end{eqnarray}

Note that $T_{D\bar{D}}/T_{equ} = 2^{-3/2}R_0/R_{10}$.
Thus in order for the $D3-\bar{D3}$ system to be at thermal equilibrium, we choose
$R_{10}=2^{-3/2} R_0$, so that $T_{D\bar{D}}= T_{equ}$.
This gives us
\begin{equation}
S_{BH} = {1\over 8}S_{D\bar{D}}
\end{equation}
With this choice, the $(-1)^{F_s}$ in the two theories are done along cycles of different asymptotic
radii. Indeed the ratio between the radii is
\begin{equation}
{\tau_{BH} \over \tau_{D\bar{D}}} = (R_0/R_{10})^{5/4} =  2^{15/8}
\end{equation}
Thus we expect the theory in the asymptotic flat regime of both sides to be different.

This suggests that the energy and temperature of the $D3-\bar{D3}$ field theory may be equal to
the mass and temperature of the $D3-\bar{D3}$ supergravity solution only up to a constant, as there is
no exact open string - closed string duality. Thus it may be possible to have $R_{10}=R_0$ and
$T_{D\bar{D}}/T_{FT}=2^{-3/2}$ with $T_{D\bar{D}}$ the temperature of the
$D3-\bar{D3}$ supergravity solution and $T_{FT}$ the temperature of the corresponding
field theory, which turns out equal to its equilibrium temperature $T_{equ}$.
In such a case the two theories (the black 3-brane and the $D3-\bar{D3}$) have the same
asymptotic flat regime (in the supergravity description), and both are in thermal equilibrium.
In fact the theories will also have the same string length, asymptotic coupling and 3-volume (the
ADM masses, however, will have a relative factor of 8; the total energy of the field theory may be
related to these by yet another factor).
Another (possibly complementary) possibility is that $n_b$,
the effective number of bosonic degrees of freedom in the $D3-\bar{D3}$ system,
is not exactly 6.

If we insist that the temperature and total energy are the same in the field theory and the supergravity
descriptions of $D3-\bar{D3}$, and we want to keep $n_b=6$,
we may still ask what would happen if $R_{10}=R_0$. This yields two theories with the same
temperature (as $\tau_{BH} = \tau_{D\bar{D}}$), asymptotic coupling, string length and
3-volume. However,
the  $D3-\bar{D3}$ system is no longer at thermal equilibrium. This means that the entropy is no longer
maximized with respect to the number of brane-antibrane pairs $N$, and the system will be unstable.
But since the tachyon between the branes and antibranes became massive with mass much greater than the
temperature, we may assume that the system is metastable; we may think of it as an $overcooled$
$D3-\bar{D3}$ system. This approximation will be valid if the time it takes for brane-antibrane pairs to
annihilate is large compared to other time scales in the problem. This scenario is further investigated
in appendix E.

\subsection{Annihilation to closed strings}

We related the chargeless black 3-brane to the $D3-\bar{D3}$ system. The descriptions of 
both systems break down at the limit $g_s N \sim 1$. It is interesting to 
see what happens to both systems as we approach this limit. We will now show 
that both behave similarly.

Let us consider at which limit the $D3-\bar{D3}$ system description breaks down.
If we adiabatically change $g_s N$  we may change the temperature accordingly in 
order to maintain (\ref{condition}) and stabilize the system.
This can no longer be done as $g_s N$ becomes of order $1$ or smaller, because 
the tachyon remains tachyonic even close to the Hagedorn temperature, and 
the brane-antibrane pairs are eventually all annihilated, with the energy emitted as
closed strings.

Let us now consider what happens to the chargeless black 3-brane at the same limit.
In the brane-antibrane system \cite{Danielsson:2001xe}
\begin{equation}
M \sim N V /g_s l_s^4
\end{equation}
This is related to a black 3-brane with mass of the same order, and this mass satisfies
\begin{equation}
M \sim r_{BH}^4 V /g_s^2 l_s^8 
\end{equation}
This implies 
\begin{equation}
r_{BH}  \sim (g_s N)^{1/4}  l_s
\end{equation}
Thus $g_s N \sim1$ corresponds to $r_{BH} \sim l_s$, which is where the black hole 
entropy becomes equal to a string entropy, and thus a transition from black hole to strings is expected
\cite{Horowitz:1996nw}. Therefore its 
energy should be emitted as closed strings for smaller $g_s N$.
We conclude that the brane-antibrane system and the black brane both annihilate into closed strings
at the same limit.

\subsection{Black hole degrees of freedom}

We related the entropies of the thermal chargeless black 3-brane and the thermal 
$D3-\bar{D3}$ system. By tracing this relation step by step
it may be possible to relate the degrees of freedom of one system 
to the degrees of freedom of the other system.
The entropies of both systems are equal only up to a numerical constant,
so perhaps only a part of the degrees of freedom can be related in this way.
By using this method, we speculate that the massless degrees of freedom of the chargeless black 3-brane,
or at least a substantial part of them, are $D3-\bar{D3}$ pairs and open strings stretched between them.

In the model \cite{Danielsson:2001xe} for the $D3-\bar{D3}$ system,
the degrees of freedom are $D3-\bar{D3}$ pairs and open strings stretched 
either between D-branes or between anti-D-branes. Let us see 
what happens to these degrees of freedom as we go through the relation depicted
in figure (\ref{fig1}). We will assume that the descent relations \cite{Sen:1999mg} among 3-branes and 4-branes
are preserved at large $g_s$, although they have been defined and proved only at $g_s=0$.

First consider a single (spectator) $D3-\bar{D3}$ pair, 
in the $D3-\bar{D3}$ background on the bottom right side of figure (\ref{fig1}).
We will now show that given the above assumption, this is related to a single (spectator) $D3-\bar{D3}$ pair on the 
chargeless black three-brane background on the bottom $left$ side of figure (\ref{fig1}).
Starting from the $D3-\bar{D3}$ system background, and moving to its S-dual background, the spectator $D3-\bar{D3}$ pair
is dual to a spectator $D3-\bar{D3}$ pair. The next step is Orbifolding by
$(-1)^{{F_s}_L}$, and this orbifold takes 
a $D3-\bar{D3}$ pair to a non-BPS D3-brane.
The orbifolded theory is type IIA string theory, and the non-BPS D3-brane 
is a kink solution of the tachyon field on a $D4-\bar{D4}$ pair. Such a pair
is wrapped over the three-torus and the time direction ($x^{10}$ in our notation),
and is infinite in one dimension. In the lift to eleven dimensions, it is described by an $M5-\bar{M5}$
pair wrapped over the three-torus, $x^{10}$ and $x^0$. 
After compactification over $x^{10}$ we get the type IIA solution on the left side of figure (\ref{fig1}),
and the spectator $M5-\bar{M5}$ pair turns into a $D4-\bar{D4}$ pair. Thus a spectator non-BPS D3-brane
on the type IIA solution of the right side of figure (\ref{fig1}) is related to a spectator non-BPS D3-brane
on the type IIA solution of the left side. In a similar manner to what we have just described, this is related to a 
single $D3-\bar{D3}$ pair on the chargeless black three-brane background on the bottom left side of figure (\ref{fig1}).

For the open strings stretched between the branes or between the antibranes in the $D3-\bar{D3}$ system, one may not 
follow the relation of figure (\ref{fig1}) so easily. Such an open string is a D1-brane in the S-dual picture, stretched between 
D3-branes or between anti-D3-branes. But after orbifolding the theory by $(-1)^{{F_s}_L}$, the D1-brane does not survive
the orbifolding, because it has a $(-1)$ charge under the $(-1)^{{F_s}_L}$ symmetry. Instead, it can be thought of as
a kink solution of the tachyon field on a single non-BPS D2-brane, but the $(-1)^{{F_s}_L}$ orbifold takes this brane to a 
BPS D2-brane, which has no tachyon field. This is because the tachyon field does not survive the orbifolding as well.

A different approach is to relate the open strings in the $D3-\bar{D3}$ system to degrees of freedom in the black three-brane
according to their relation to the spectator $D3-\bar{D3}$ pairs.
Since spectator $D3-\bar{D3}$ pairs in the $D3-\bar{D3}$ system are related to spectator $D3-\bar{D3}$ pairs in the 
chargeless black three-brane, it is most natural to assume that open strings stretched between the branes (antibranes)
in the $D3-\bar{D3}$ system are related to open strings stretched between the branes (antibranes) in the 
chargeless black three-brane.

\subsection{Charged black holes}

We will now turn to discuss the relation between a charged black three-brane and a charged $D3-\bar{D3}$ system, 
for a small charge, still far from extremality. Note that the $D3-\bar{D3}$ system must
have unequal number of branes and antibranes.
From the field theory point of view \cite{Danielsson:2001xe} it turns out that the 
entropies of both systems are equal, up to the same numerical constant as in the uncharged case, 
if one assumes that in the $D3-\bar{D3}$ system
the two gasses of massless open strings, one living on the D-branes 
and the other living on the anti-D-branes, have the same energy density (or pressure) rather
than the same temperature. This is possible since they are effectively decoupled from each other. 
The same assumption in the $D3-\bar{D3}$ field theory must be made in order to reproduce the black hole
low-frequency absorption and emission probabilities, again up to numerical factors 
\cite{Garcia:2004vx}.
For an alternative field theory approach see appendix F.

We would like to relate the charged black three-brane and the charged $D3-\bar{D3}$ system in the supergravity framework, 
as we did for the chargeless case. However a simple generalization is not possible, as will be explained 
below. Instead, for small charges (far from extremality) we suggest the following construction of the relation:
We have seen in the previous subsection that a single $D3-\bar{D3}$ pair in the chargeless $D3-\bar{D3}$ system is related to 
a single $D3-\bar{D3}$ pair in the chargeless black three-brane. This means that a single D3-brane in each system is related 
to a single D3-brane 
in the other. Adding many D3-branes to both systems would make them charged, and we get a relation between 
a charged black three-brane and a charged $D3-\bar{D3}$ system.
However this does not explain the condition mentioned above, needed for the field theory description to give 
the black hole entropy correctly, namely that the open string gas on the 
D-branes and the open string gas on the 
anti-D-branes must have the same energy density.

A simpler generalization of our method to the charged case is not possible,
because the four-form is not invariant under the $(-1)^{{F_s}_L}$ symmetry, so this
orbifold cannot be defined on such a background.
Using T-duality instead, in order to move from a type IIB solution to a type IIA solution, 
breaks down the supergravity approximation, in particular because
the 3-volume becomes very small (compared to the string length) after T-duality.

\section{Discussion and generalizations}

We described a relation between the thermal chargeless black 3-brane and
the thermal version of a system that we interpret as the $D3-\bar{D3}$ system,
for a particular value of the asymptotic string coupling $g_s$ (with arbitrary string length $l_s$ and
black brane mass; alternatively, $g_s$ and the mass are arbitrary and the relation holds only for
a particular value of $l_s$). In particular we expect their entropies to be of the same order of magnitude. This may explain the
comparison between the Bekenstein-Hawking entropy of the black 3-brane and the field theory
entropy of the thermal $D3-\bar{D3}$, which has been studied in \cite{Danielsson:2001xe}. The
relation that we found agrees with that comparison, up to some numerical factors in the different
quantities in the problem (temperature, mass and entropy).
These factors may be the result of moving from the closed string picture to the open string picture though
there is no complete open-closed string duality; Thus the temperature, mass etc. may not be exactly the same
in the two pictures.

It is important to note that the approximation of \cite{Danielsson:2001xe}
that the massless degrees of freedom on the branes and
antibranes are decoupled from each other, which was the basis for their field theory description of the thermal
$D3-\bar{D3}$, is arbitrarily good when $l_s$, $g_s$ are arbitrarily
small and $g_s N$ arbitrarily large, but only for a specific temperature. Thus, the system is stable as
long as the temperature is fixed at the appropriate value, as we are doing when we study the theory on a
Euclidean time circle and set its radius. However, the system is unstable once the temperature is allowed
to change (it has a negative specific heat \cite{Danielsson:2001xe}, like the black 3-brane); Therefore a
duality between open and closed strings holds only for a specific energy scale, which agrees with the
absence of a decoupling limit between open and closed strings for the corresponding geometry.


One may wonder what would happen for other values of $g_s$. The relation that we have shown
cannot be simply generalized because $g_s$ of one solution (either the black brane or the $D-\bar{D}$)
translates to a combination of $g_s$ and the temperature in the other, and each solution is stable only for
a specific value of the temperature.
One possibility is that the relation is still valid for other values of $g_s$. The Bekenstein-Hawking
formula for black hole entropy depends on $g_s l_s^4$ rather than on $g_s$ and $l_s$ independently,
so it is conceivable that a black 3-brane with particular $g_s$ and $l_s$ can be described by a
black 3-brane with different values of $g_s$ and $l_s$, as long as $g_s l_s^4$ is the same. If this is
true, then any thermal black 3-brane is related to a thermal $D3-\bar{D3}$, where the
latter has the appropriate relation between $g_s$ and $l_s$.
Another possibility is that the black 3-brane and the $D3-\bar{D3}$ system are two different phases
in the space of solutions, and the phase transition occurs near the point that we have identified. If the
phase transition is of the 2nd order this would explain why the two systems are similar at this point.

Our analysis cannot be extended simply to $p \ne 3$. For $p=1,5$, the S-dual of a $Dp-\bar{Dp}$ 
system consists of fundamental strings or NS branes. If these are left unchanged by the 
$(-1)^{{F_s}_L}$ orbifold, as they are for $g_s=0$, then the description of the system in type IIA may be
different than the one we have found. For an even $p$ one
should use type IIA superstring, but then one cannot perform the final S-duality we have used, and the
asymptotic string coupling remains large ($g_s \gg 1$).
A field theory description in terms of a $Dp-\bar{Dp}$ system for $p<6$, in strong coupling ($g_s N \gg 1$),
has been given in \cite{Danielsson:2001xe,Saremi:2004pi,Bergman:2004tz}, including for a non-vanishing angular momentum,
with a similar success as in the $p=3$ case. A different counting, in terms of branes and fundamental strings, have 
been suggested in \cite{Horowitz:1996ay} for $p=1,5$. The interested reader may refer to
\cite{Rabadan:2002zq,Dabholkar:1994ai} which have used a different method (allowing a conical
singularity) for $p=6$ and have compared either the black brane or the
brane-antibrane system to an orbifold construction in the supergravity framework; however
one should bear in mind that in such a case $R_{10}\ll (M_{11})^{-1}$ if $g_s$ is small, since
$g_s = (R_{10}M_{11})^{3/2}$, which makes the 11 dimensional
supergravity approximation dubious.

\section*{Acknowledgements}

We would like to thank N. Itzhaki, B. Kol, G. Mandal and D. Reichmann for useful discussions.
This work is supported by DIP and the ISF excellence center.

\appendix

\section{The energy-momentum tensor at the singularity}

In this section we compute the components of the energy-momentum tensor at the singularity in $r=r_0$
by the method of \cite{Das:1997tk}, for all the chargeless asymptotically flat
supergravity solutions with
symmetry $ISO(p+1)\times SO(9-p)$, i.e. (\ref{Poincaremetric}) with $c_2=0$. We show that they all
obey the relation $T_{ij}= - T_{00}\delta_{ij}$, which was suggested in \cite{Danielsson:2001xe} to
hold for the brane-antibrane system at zero temperature.
A similar computation has been done in \cite{Ohta:2002ac}, with a different interpretation.

The asymptotic Einstein frame metric, to sub-leading term in ${1\over (r-r_0)^{7-p}}$, is
$g_{\mu\nu}= \eta_{\mu\nu} + h_{\mu\nu}$ with
\begin{eqnarray}
h_{00} &=&   {7-p\over 16} \left((3-p)c_1 -4 c_3 k  \right)  \left({r_0\over r-r_0}\right)^{7-p} \nn
h_{ij} &=& - {7-p\over 16} \left( (3-p)c_1 -4 c_3 k  \right)  \left({r_0\over r-r_0}\right)^{7-p}
\delta_{ij}\nn
h_{ab} &=& {p+1\over 16} \left((3-p)c_1 -4 c_3 k  \right)  \left({r_0\over r-r_0}\right)^{7-p}
\delta_{ab}
\end{eqnarray}
where $i$, $j$ stand for the directions parallel to the brane, and $a$, $b$ for the directions
perpendicular to it.

$h_{\mu\nu}$ satisfies the harmonic gauge condition
\begin{eqnarray}
\partial_\lambda {h_\mu}^\lambda - {1\over 2} \partial_\mu h = 0 &,&
h \equiv \eta^{\mu\nu} h_{\mu\nu}
\end{eqnarray}

The linear Einstein equations simplify to
\begin{equation}
\partial_\lambda \partial^\lambda \left( h_{\mu\nu} -{1\over 2} \eta_{\mu\nu} h\right) =
-16 \pi G T_{\mu\nu}
\end{equation}

This gives
\begin{eqnarray}
T_{00} &=& {7-p \over 32\pi G}\omega_{8-p} \left((3-p)c_1 -4 c_3 k  \right)
{r_0}^{7-p} \delta^{9-p} ( x_\perp ) \nn
T_{ij} &=& -{7-p \over 32\pi G} \omega_{8-p}\left((3-p)c_1 -4 c_3 k  \right)
{r_0}^{7-p} \delta^{9-p} ( x_\perp ) \delta_{ij}\nn
T_{ab} &=& 0
\end{eqnarray}
where $\omega_{8-p}$ is the $8-p$ volume of an $S^{8-p}$ of unit radius, 
and $\delta^{9-p} ( x_\perp )$ is a delta function in all directions perpendicular to the brane.

\section{The horizon 8-volume and temperature}

In a part of the space of the asymptotically flat solutions described in \cite{Zhou:1999nm}
(which includes all solutions of the type (\ref{Poincaremetric})), $g_{tt}$ vanishes as
$r\rightarrow r_0$.
We will refer to this limit as a "horizon", although this is not precise since (except for in the
Strominger-Horowitz black brane) the curvature is singular at this limit, so geodesies cannot be
continued beyond this point.

The "horizon" 8-volume, for $0 \le p < 7$, is given by
\begin{eqnarray}
A_H = \int_{r=r_0} \prod_{i=1}^{8-p} \sqrt{g_{ii}}d\theta_i
\prod_{\alpha=1}^p \sqrt{g_{\alpha\alpha}} dx^\alpha
&=& \lim_{r\rightarrow^+ r_0} \omega_{8-p} V_p {r_0}^{8-p} \, \cdot e^{(8-p)B+p A}
\sim \lim_{r\rightarrow^+ r_0}{f_-}^W (r) \nn
\end{eqnarray}
Here $i$ runs over angles in $S^{8-p}$ and $\alpha$ over vectors in $T^p$.
$V_p$ is the $T^p$ volume and $\omega_{8-p}$ is a unit $S^{8-p}$ volume.
$e^{2A} = g_{\alpha\alpha}$ ($\alpha=1...p$) and $e^{2B} = g_{rr}$ in the Einstein frame, in 
isotropic coordinates.
$f_-$ is defined in (\ref{Poincaremetric}).

W is defined as
\begin{equation}
W \equiv1+{1\over 7-p} + {p-3\over 8}c_1+{p+9\over 32}c_2
+{1\over 2} c_3 k
\label{W}
\end{equation}
with $c_{1,2,3}$ and $k$ as in (\ref{Poincaremetric}).
This can be generalized for the charged case (i.e. general $c_3$) by replacing $c_3\rightarrow -1$
for every $c_3 \ne 1$. For $p=-1$, $W$ is not given by (\ref{W}), but is rather equal to ${9\over 4}$.

It can be shown that $W$ is always positive, which implies that the
"horizon" 8-volume vanishes, except for the black brane solution (\ref{regsol}), where $W=0$ and the 
horizon area is finite.

The temperature $T$ of a geometry with a horizon can be computed in two different methods, both
yielding the same value:\\
(a) By performing a Bogoliubov transformation between propagating modes locally defined in past
null infinity and propagating modes locally defined in future null infinity. This gives particle production
rates which correspond to thermal radiation of temperature $T=\kappa/2 \pi$ with $\kappa$ the
horizon surface gravity.\\
(b) By analytically continuing to Euclidean metric ($t\rightarrow i x^0$). There is no conical singularity
only if the periodicity of $x^0$ is precisely $\beta\equiv 2\pi/\kappa$.

We may want to interpret the limit $r\rightarrow r_0$ as a horizon, and calculate the temperature.
Both methods yield an infinite temperature for all solutions except for the
Strominger-Horowitz black branes. This result agrees with the
thermodynamic relation
\begin{equation}
\beta\equiv {1\over T} =
{\partial S\over \partial U}
= \hbox{const.}\cdot{\partial A_H \over \partial M}
\label{thermo}
\end{equation}
where $A_H$ is the horizon area (or 8-volume in the 10 dimensional case) computed in the Einstein 
frame. In our case $A_H$ is zero while $M$ depends on the parameters $c_1,c_2,r_0$, so
the right hand side of (\ref{thermo}) is zero. This indeed implies $T\sim\infty$.

The temperature is calculated as follows:
We start with the metric
\begin{equation}
\dd s^2 = - F_1 (r) \dd t^2 + F_2(r) \dd r^2 + ...
\end{equation}
where $F_1$ vanishes for $r\rightarrow r_0$.
We make the coordinate change $r \rightarrow \rho$ where
\begin{equation}
\rho = \int_{r_0}^r \sqrt{F_2(\tilde{r})} \dd \tilde{r}
\end{equation}
Then
\begin{equation}
\dd s^2 = - F_1 \left(r(\rho)\right) \dd t^2 + \dd \rho^2 + ...
\end{equation}
with $F_1(r)$ vanishing at $\rho = 0$.
If near that point $F_1(r) \sim \kappa^2 \rho^2$ for some constant $\kappa$ then the temperature
is $T=1/\beta = {\kappa\over 2\pi}$. Thus
\begin{equation}
2\pi T = \kappa = {\dd \sqrt{F_1} \over \dd \rho}_{|\rho =0}
= \lim_{r \rightarrow r_0}{{F_1}^\prime \over 2\sqrt{F_1 F_2}}
\label{kappa}
\end{equation}
with the prime denoting a derivative with respect to $r$.
The surface gravity at $r=r_0$ is equal to $\kappa$.

For the solutions to the metric we are interested in, $F_1$ and $F_2$ are given by
(\ref{Poincaremetric})
\begin{eqnarray}
F_1 &=& F^{\gamma_1} \nn
F_2 &=& \left({f_- \, f_+}\right)^{ 2\over 7-p  }
F^{\gamma_2} \nn
\gamma_1 &\equiv& { (7-p) (3-p) \, c_1 + (2(7-p) +{1\over 4} (3-p)^2 - 32) c_2
-  4 (7-p) c_3 k \over  32 } \nn
\gamma_2 &\equiv& {- (p+1)(3-p)\, c_1   +  {1\over 4} (3-p)^2 c_2
+ 4 (p+1) c_3 k \over 32 } \nn
F &\equiv& {f_- \over f_+}\nn
f_\pm &\equiv& 1 \pm \left( r_0 \over r \right)^{7-p}\\
\end{eqnarray}
$F$ and $f_-$ vanish as $r\rightarrow r_0$.
We are interested in the case $\gamma_1 > 0$, so that $F_1$ vanishes at this limit as well.
(\ref{kappa}) is
\begin{eqnarray}
\lim_{r \rightarrow r_0} {{F_1}^\prime \over 2\sqrt{F_1 F_2}}  =
\lim_{r \rightarrow r_0} {\gamma_1\over 2}
F^\prime \left({f_- \over f_+}\right) ^{(\gamma_1 - \gamma_2)/2 -1}
\left({f_- \, f_+}\right)^{-{1 \over 7-p }}
\end{eqnarray}
$F^\prime (r_0) = {7-p \over 2 r_0}$ and $f_+ (r_0) = 2$ . Therefore the temperature is
\begin{eqnarray}
T &=& {\kappa \over 2 \pi} = 2^{-\gamma -{2 \over 7-p }-3} \,
{7-p \over \pi r_0} \gamma_1
\lim_{r \rightarrow r_0} \cdot {f_-}^\gamma \nn
\gamma & \equiv & {(\gamma_1 - \gamma_2)/2 -1 -{1 \over 7-p }}
\end{eqnarray}
T is finite for $\gamma=0$. However this happens only for the Strominger-Horowitz black brane
solution (\ref{regsol}). For all other solutions (\ref{Poincaremetric}) 
$\gamma<0$ and the temperature is infinite. Note that $\gamma = -W$ with $W$ defined in (\ref{W}).
Here, too, the results can be extended to the charged case, by replacing $c_3\rightarrow -1$ for every 
$c_3 \ne 1$.

Note that this calculation is done in Einstein frame, where the equations of motion take a simple
form\footnote{Incidentally, if one replaces the Einstein frame metric by the string frame metric, the result
does not change, again yielding an infinite temperature whenever $g_{tt}$ vanishes at
$r\rightarrow r_0$, because shifting from the Einstein frame to the
string frame is implemented by shifting $\gamma_1$ and $\gamma_2$ by an equal constant and so
$\gamma$ remains unchanged; We assume $g_{tt}$ still vanishes at $r\rightarrow r_0$ so $\gamma_1$ 
is positive even after the shifting. Nevertheless, we do not know if this result
has any meaning.} \footnote{This has nothing to do with the fact that the proper normalization of the
temperature, when comparing supergravity and field theory as in
\cite{Itzhaki:1998dd,Maldacena:1997cg}, is such that the length units are normalized so that the string
frame metric is asymptotically $\eta_{\mu\nu}$ (rather than the Einstein frame metric).}.

Upon lifting to 11 dimensions, we may perform the same procedure in
the 11 dimensional metric, with the only change being a shift of $\gamma_1$ and $\gamma_2$ by an
equal constant, so that $\gamma$ remains unchanged. Thus the Hawking temperature in 11 dimensions
is still infinite, except for when the new $\gamma_1$ (after the shift) is zero, in which case the
time direction is flat and any temperature can be defined. For solutions (\ref{Poincaremetric}) with the
symmetry $ISO(p+1)\times SO(9-p)$, namely $c_2=0$, this happens precisely for the case
(\ref{goodsing}), and in addition for the same case but with the sign of $c_1$ and $c_3$ flipped
(namely, the case $(c_1,c_2,c_3)=(-{12-2p \over 7-p},0,sign(p-2))$); Equivalently, we take
$r_0\rightarrow -r_0$. It is easy to verify that the last
set of solutions have a negative ADM mass in 10 dimensions.

The infinite temperature can be understood as follows: For a sub-extremal black brane,
the temperature is positive and the singularity is hidden behind the horizon, which
is a surface of infinite redshift. For a super-extremal black
brane, the temperature is negative and the singularity is naked, with the "horizon"
(a surface of infinite redshift) behind it. In our case the singularity is infinitely redshifted and so
lies "on the horizon", in a sense, and we get an infinite temperature. In the extremal case the
temperature is zero.

Finally we give the $D6-\bar{D6}$ as an explicit example, in light of the second approach
mentioned above, namely computing the temperature via the analytic continuation to Euclidean
metric and curing its conical singularity by making $t$ periodic.

The Einstein frame metric of the $D6-\bar{D6}$ in isotropic coordinates is 
\cite{Brax:2000cf,Sen:1997pr}
\begin{equation}
\dd s^2 = \left({r - r_0 \over r + r_0}\right)^{1/4} \dd x_\mu \dd x^\mu  +
r^{-4} (r - r_0)^{1/4} (r + r_0)^{15/4} \left(\dd r^2 + \dd {\Omega_2}^2 \right)
\end{equation}
with $\mu = 0,1... 6$.
For $r\sim r_0$ let us define $\hat{r} = (r-r_0)^{9/8}$. Then at this limit, suppressing
factors of $r_0$ and $2$ (which can be omitted by a rescaling of the coordinates), we obtain
\begin{equation}
\dd s^2 \sim -{\hat{r}}^{2/9} \dd t^2 + \dd {\hat{r}}^2 + ...
\end{equation}

Taking only this part of the metric and analytically continuing to a Euclidean metric
$t\rightarrow i x_0$ gives
\begin{equation}
\dd {s_E}^2 = {\hat{r}}^{2/9} \dd {x_0}^2 + \dd {\hat{r}}^2
\end{equation}
This metric has a singularity at $\hat{r} = 0$ which is not conical (in fact, a naked curvature
singularity). Clearly it cannot be cured by making $x_0$ periodic. Thus no finite temperature can be
naturally assumed.

\section{The effective potential for a minimally coupled scalar}

Dp-branes for $p<6$ admit a decoupling limit \cite{Maldacena:1997re}, for which the gravitational
source (i.e. the brane) is decoupled from the asymptotic regime. This limit can be implemented either
by taking a near-horizon limit of the metric, or by computing the scattering of gravitons and minimally
coupled scalars off the D-brane and taking the limit where the former decouple from the latter
\cite{Alishahiha:2000qf}. The last method can be
qualitatively appreciated also by computing the effective potentials felt by a minimally coupled scalar
and by a graviton; in the decoupling limit the potential will have a barrier of infinite height which scatters
the scalar and the graviton back. By this method it has been shown in \cite{Alishahiha:2000qf}
that D6-branes have no decoupling limit.

In this section we show that there is no decoupling limit in any of the chargeless solutions,
by calculating the effective potential felt by a minimally coupled scalar
and show that there is no limit in which the potential has an infinitely-height barrier. In fact, the potential is
always negative. This is an extension of results appearing in \cite{Brax:2000cf}.
Explicit calculations of scattering amplitudes are beyond the scope of this work.

A minimally coupled scalar $\Phi(r,t) = \Phi(r)e^{i\omega t}$ admits the following equation of motion
\begin{equation}
0 = \nabla_\mu \nabla^\mu \Phi = g^{-{1\over 2}}\partial_\mu \left( g^{1\over2} \partial^\mu \Phi\right) =
\left[-g^{tt}\omega^2+g^{-{1\over 2}}\partial_r\left(g^{1\over 2}g^{rr}\partial_r\right)\right]\Phi
\end{equation}
where the metric is in the Einstein frame, is assumed to depend only on $r$, and $g$ is its determinant.

For a metric with $ISO(p)\times SO(9-p)$ symmetry,
using the notations of \cite{Zhou:1999nm} we get the following equation for $\Phi(r)$
\begin{equation}
\Phi^{\prime\prime}-{h^{\prime\prime}\over h^\prime}\Phi^\prime
+ {e^{2(B-A)}\over f}\omega^2\Phi = 0
\end{equation}
in isotropic coordinates, where $-f e^{2A} = g_{tt}$, $e^{2A} = g_{\alpha\alpha}$ ($\alpha=1...p$),
$e^{2B} = g_{rr}$ and $h=\ln (f_-/f_+)$ where $f_\pm$ defined as in (\ref{Poincaremetric}).
A prime stands for a derivative with respect to $r$.

By redefining the scalar field: $\varphi(r) \equiv \Phi(r) {h^\prime}^{-1/2}$ we get the Schr\"odinger
equation:
\begin{equation}
\varphi^{\prime\prime} - V(r) \varphi = 0
\end{equation}
with
\begin{equation}
V(r) = {1\over 4} \left({h^{\prime\prime}\over h^\prime}\right)^2
-{1\over 2}\left({h^{\prime\prime}\over h^\prime}\right)^\prime
-\omega^2 {e^{2(B-A)}\over f}
\end{equation}

For chargeless solutions ($c_3=\pm1$) we get
\begin{eqnarray}
V(r) &=&
-{1\over r^2}\left[{(7-p)^2\left({r_0\over r}\right)^{2(7-p)}\over
\left[1-\left({r_0\over r}\right)^{2(7-p)}\right]^2}-{1\over 4}(7-p)^2 + {1\over 4}\right] \nn
&-& \omega^2\left[1-\left({r_0\over r}\right)^{2(7-p)}\right]^{2\over 7-p}
\cdot\left[{1-\left({r_0\over r}\right)^{7-p} \over 1+\left({r_0\over r}\right)^{7-p}}\right]^\beta  \nn
\beta &\equiv& {p-3\over 4} c_1+{p+9 \over 16}c_2 +c_3 k
\end{eqnarray}
with $c_{1,2,3}$ and $k$ as in (\ref{Poincaremetric}).

$V$ is always negative for $p=6$. For a lower $p$ and for low enough $\omega$, $V$ is positive in
some range of $r$ and has a maximum, but there is no limit in which this maximum becomes infinite.
Therefore there is no limit in which the scalar is decoupled from the gravitational source.

Note that for the black brane solutions (\ref{regsol}) , $\beta = -2 - {2\over 7-p}$, while for
the solutions (\ref{goodsing}), which we interpret as the brane-antibrane system,
$\beta = - {2\over 7-p}$.

As examples, we give here the potential $V$ for these solutions with $p=3$ and $p=6$,
in isotropic coordinates (we take units in which $r_0=1$ for simplicity)
\begin{eqnarray}
V_{black~6-brane} &=& -{1\over (1-r^2)^2} - \omega^2{(1+r)^6\over (1-r)^2 r^4}\nn
V_{black~3-brane} &=& {15 - 94 r^8 + 15 r^{16}\over 4(1-r^8)^2 r^2}
- \omega^2{(1+r^4)^3\over (1-r^4)^2 r^4}\nn
V_{D6 \bar{D6}} &=& -{1\over (1-r^2)^2} - \omega^2{(1+r)^4\over r^4} \nn
V_{D3 \bar{D3}} &=& {15 - 94 r^8 + 15 r^{16}\over 4(1-r^8)^2 r^2}
- \omega^2{1+r^4\over r^4}
\end{eqnarray}

\section{Comments on orientation of non-BPS branes}

If non-BPS branes have a relative orientation, so that
two non-BPS branes can have either the same orientation or opposite orientations, then the 
system of $N$ non-BPS branes that is discussed in section 3
may be interpreted as consisting of $N_1$ non-BPS branes of one
orientation and $N_2$ of the opposite orientation. Thus its orbifold results in $N_1$ branes and
$N_2$ antibranes.

It has been shown \cite{Sen:1998ex,Sen:1999mg} that the non-BPS Dp-brane in type IIB (IIA) can
also be thought of as a kink solution of the complex tachyon on a $D(p+1) - \bar{D}(p+1)$ pair of
type IIB (IIA).
Similarly, the BPS D(p-1)-brane (antibrane) is a kink (anti-kink) solution of the real tachyon on a
non-BPS Dp-brane.

It has been shown in \cite{Sen:1998ex}, for a particular example, that a $D(p-1)$-brane and a 
$D(p-1)$-antibrane far away from each other, on two apposite points of a circle, can be 
described by a kink and an anti-kink solutions of  the real tachyon on a non-BPS Dp-brane, glued to 
each other.

Similarly, for the complex tachyon field on a $D(p+1) - \bar{D}(p+1)$ pair on a circle, we expect there
to be a solution which can be described as a kink - anti-kink pair, at least for a large enough circle.
We later show that every complex tachyon potential $V(|T|)$ which has a saddle point
at $|T|=T_0$, $T_0\ne 0$, indeed has a 
solution on a circle which describes a kink-antikink pair as the circle circumference approaches infinity.
We find the solution explicitly for a potential of the simplified form
$V(|T|) = -{m\over 2}T^2 + {\lambda \over 4} T^4$.

These kink and anti-kink have opposite orientations in spacetime (they are related by a $Z_2$ reflection 
in the $x^{p+1}$-axis). Thus they describe two non-BPS $Dp$-branes which are not identical,
which we shall call two non-BPS $Dp$-branes of opposite orientations.
By generalization from the $D(p-1) - \bar{D(p-1)}$ case, it should be expected that two non-BPS
$Dp$-branes of opposite orientation can also be adjacent.

This does not mean that non-BPS branes have a $Z_2$ conserved charge.
Because a non-BPS $Dp$-brane is a kink solution of a complex tachyon field, there is a continuous
deformation between the kink and the anti-kink solutions, i.e. between the two non-BPS branes
discussed above. Both can also be deformed to the vacuum, and indeed a
non-BPS brane of this type is unstable.
This is in contrast to the $Dp$-brane and the antibrane, which cannot be
continuously deformed to each other and have in fact different charges under the RR $p+1$-form.

Let us first show an explicit example with
\begin{equation}
V(|T|) = -{m^2 \over 2}|T|^2 + {\lambda \over 4} |T|^4
\end{equation}
Let us consider solutions which depend only on $x^{p+1}$. For simplicity, we denote 
$x\equiv x^{p+1}$.

The equation of motion reads
\begin{equation}
{\partial^2 T \over \partial x^2}= -m^2 T + \lambda T |T|^2
\end{equation}
By rescaling $x$ and $T$ we may fix $m=1$, $\lambda=1$.
Then the most general solution for which $T=0$ at $x=0$ is given by the Jacobi elliptic function
\begin{equation}
T = e^{i\theta} \sqrt{1-a} \, \cdot \, sn\left(\sqrt{1+ a \over 2} x \bigg| {1-a \over 1+a}\right)
\end{equation}
with $a \equiv \sqrt {1-2 |T'(0)|^2}$ and $\theta$ is an overall phase.
For $|T'(0)|=1/\sqrt{2}$ this is the kink solution $tanh({x\over \sqrt{2}})$. For smaller  $|T'(0)|$ the 
solution is sine-like periodic, and for a larger $|T'(0)|$ the solution is tangent-like. For 
$|T'(0)|<{1\over 2}$ the periodicity is
monotonically increasing in $|T'(0)|$. Thus on a large circle, there is
a solution with precisely one period, which resembles a kink-antikink solution, and at the limit where the
size of the circle goes to infinity, this solution can be seen as gluing a kink and an anti-kink solutions.

Finally, we generalize this result to any $V(|T|)$ which has a first saddle point at $|T|=T_0$,
$T_0\ne 0$.
The equation of motion is
\begin{equation}
{\partial^2 T \over \partial x^2}= {dV(|T|) \over d\bar{T}} = 2 F'(|T|^2) T
\end{equation}
where $F(|T|^2) = V(|T|)$.
Both sides have the same phase. For a solution which satisfies $T(0)=0$, the solution depends only on
$\partial_x T_{|x=0}$. Thus the solution has an overall phase which is equal to that of 
$\partial_x T_{|x=0}$, and otherwise it is enough to solve for real $T$.

The most general non-trivial solution with real $T$ and $T(0)=0$ is
\begin{equation}
x (T) = \pm \int {dT \over \sqrt {2 \left(V(|T|)-V(T_0)\right) + c}}
\label{xT}
\end{equation}
note that this function has an odd parity.

For $c=0$ the integrand diverges at $T=T_0$. Thus this is the kink (or anti-kink) solution,
with $|T|\rightarrow T_0$ at $x\rightarrow \pm \infty$ (that $x$ reaches infinity as $|T|\rightarrow T_0$
can bee seen by noting that otherwise we would have an $x$ at which both $dT/dx$ and $d^2T/dx^2$
vanish, and the $T$ would therefore be constant).

For a positive $c$, the integrand is finite even at $T=T_0$ and the solution $T(x)$ goes beyond the
$T_0$ point, as in the tangent-like solutions described above for the polynomial potential.
If $T_0=\infty$, however, this is just another kink (or anti-kink) solution.

For a negative $c$, the integrand diverges at $T=T_1$ for some $T_1<T_0$. The integral from
$0$ to $T_1$ converges (this is because $dV/dT$ is non-zero at $T_1$, so the contribution to the
integral near $T_1$ is $\sim \int {dT\over \sqrt{T-T_1}}$). Let us denote it $x_1\equiv x(T_1)$.
$dx/dT$ diverges at $T_1$; Thus $dT/dx$ vanishes at $\pm x_1$. We will now show that the solution
$T(x)$ is a sine-like periodic function.\\
The full solution $T(x)$ is an extension of the inverse function of (\ref{xT}) beyond the range
$(-x_1,x_1)$.
Through the equation of motion, the second derivative of $T(x)$ is equal to a function of
$T$ only. Thus every even derivative of $T$ with respect to $x$ is equal to a sum of terms of the form 
$f_k(T) (\partial_x T)^{2k}$,
and every odd derivative of $T$ with respect to $x$ is equal to a sum of terms of the form 
$\hat{f}_k(T) (\partial_x T)^{2k+1}$,
with $k \ge 0$ and $f_k(T)$, $\hat{f}_k(T)$ some functions of $T$ only.
$\partial_x T_{|x=x_1}=0$, so all the odd derivatives of $T$ vanish at $x_1$. Thus $T(x)$ can be
smoothly continued beyond $x_1$ according to $T(x) = T(2 x_1-x)$. Similarly, $T$ can be continued
beyond $-x_1$. $T$ is therefore a function of odd parity, extrema at $\pm x_1$ and periodicity
$4 x_1$. This solution can be put on a circle of this circumference.

To conclude, every potential $V(|T|)$ with a first saddle point at $|T|=T_0$, $T_0\ne 0$, has a
kink solution. On a finite circle it has a solution for which $T$ is always finite and crosses zero once in
each direction. As the circumference of the circle approaches infinity, this solution describes a kink -
antikink pair.

\section{An overcooled $D3-\bar{D3}$ system}

Suppose that we insist that the temperature and total energy are the same in the field theory and the
supergravity descriptions of $D3-\bar{D3}$, and we want to keep $n_b=6$.
Let us see what would happen if $R_{10}=R_0$. This yields two theories with the same
temperature (as $\tau_{BH} = \tau_{D\bar{D}}$), asymptotic coupling, string length and
3-volume. However,
the  $D3-\bar{D3}$ system is no longer at thermal equilibrium. This means that the entropy is no longer
maximized with respect to the number of brane-antibrane pairs $N$, and the system will be unstable.
But since the branes and antibranes are decoupled\footnote{in particular, the tachyon - which is related
with the annihilation of  brane-antibrane pairs - is massive with mass much greater than the
temperature.}, we may assume that the system is metastable; we may think of it as an overcooled
$D3-\bar{D3}$ system. This approximation will be valid if the time it takes for brane-antibrane pairs to
annihilate is large compared to other time scales in the problem.

The mass and entropy of the $D3-\bar{D3}$ system can be written in terms of
$N$ and the temperature $T$ as \cite{Danielsson:2001xe}
\begin{eqnarray}
M_{D\bar{D}} &=& 2N \tau_3 V + {3\over 4} \pi^2 N^2 V T^4
\nn
S_{D\bar{D}}&=& \pi^2 N^2 V T^3
\end{eqnarray}
where $\tau_3 \equiv 1/(2\pi)^3 g_s {l_s}^4$ denotes the D3-brane tension, and
we are suppressing the ${D\bar{D}}$ subscript for convenience when there is no ambiguity.

For a given temperature $T$ and 3-volume $V$ we get
\begin{eqnarray}
S_{D\bar{D}} &=& {4 \left(\sqrt{3 M \pi ^2 V T^4+4 V^2 {\tau_3} ^2}-2 V
   {\tau_3} \right)^2 \over 9 \pi ^2 T^5 V}
\end{eqnarray}

Since in this scheme
\begin{equation}
T_{D\bar{D}} = T_{BH} = 1/\pi r_{BH} =
1/\pi r_{D\bar{D}} =
5^{1/4}2^{-5/2} \pi^{-2} {g_s}^{-1/2} {l_s}^{-2} (M_{D\bar{D}}/V)^{-1/4}
\end{equation}
The entropy is
\begin{eqnarray}
S_{D\bar{D}} &=& {143-16 \sqrt{79}\over 9} 2^{9\over 2} 5^{-{5\over 4}}\pi ^2
{g_s}^{1/2} {l_s}^2 M_{D\bar{D}}^{5/4} V^{-{1\over 4}}
\end{eqnarray}
Which has the same functional form as (\ref{SBH}) and (\ref{SDDbar}), up to a numerical factor.

By substituting for the values found in (\ref{TglDDbar}, \ref{MassDDbar})  with $R_{10}=R_0$ we
get
\begin{eqnarray}
S_{D\bar{D}} &=& S = {143-16 \sqrt{79} \over 9} {M_{11}}^{13}
{R_0}^{10} \prod_{i=1}^3 R_i = {143-16 \sqrt{79} \over 72} S_{BH}\sim 10^{-2}S_{BH}
\end{eqnarray}

\section{Introducing a chemical potential to the $D3-\bar{D3}$ FT}

A charged $D3-\bar{D3}$ system whose field theory reproduces the charged black three-brane entropy,
up to a numerical factor, has been given in \cite{Danielsson:2001xe}. We explore here 
another possibility, which is to introduce a chemical potential to the $D3-\bar{D3}$ system \footnote{We thank Z. 
Komargodski for this suggestion.}. Thus the energy and entropy of the system will be
\begin{eqnarray}
M &=& (N + \bar{N}) \tau_3 V + n_b {\pi^2\over 16} (N^2 + \bar{N}^2) V T^4
+ (N-\bar{N}) \mu
\nn
S &=& n_b {\pi^2 \over 12} (N^2 + \bar{N}^2)  V T^3
\end{eqnarray}
where $N$ and $\bar{N}$ are the numbers of branes and anti-branes, respectively, $\tau_3$ is their
tension, and $\mu$ is the
energy cost of having more branes than anti-branes (i.e. of having a non-vanishing net charge).
$S$ as a function of $M$, $N$, $\bar{N}$, $V$ and $\mu$ is
\begin{equation}
S = {2\over 3} \pi^{1/2} \left((N^2 + \bar{N}^2) n_b V \right)^{1/4}
\left(M -  (N-\bar{N}) \mu -(N + \bar{N}) \tau_3 V \right)^{3/4}
\end{equation}
minimizing with respect to $N$ and $\bar{N}$ we get $S$ as a function of $M$, $V$ and $\mu$,
and
\begin{equation}
\mu = {-M + \sqrt{M^2 - 25 V^2 (N-\bar{N})^2 {\tau_3}^2} \over 5  (N-\bar{N})}
\end{equation}

Replacing $\mu$ for this expression in $S(M,V,\mu)$ and using $Q=N-\bar{N}$
we get $S(M,V,Q)$. We are interested in the far-from extremal regime and we thus expand around
$Q=0$ and get (assuming $n_b=6$)
\begin{equation}
S = 2^{3/2} 5^{-5/4}\pi^{1/4} \kappa^{1/2} V^{-1/4} M^{5/4}
- 2^{-5/2}5^{3/4} \pi^{5/4} \kappa^{-3/2} V^{7/4} M^{-3/4} Q^2 +O\left(Q^{4}\right)
\label{chargedFT}
\end{equation}
with $\kappa = \sqrt{\pi}/\tau_3$.

The supergravity entropy\footnote{see \cite{Danielsson:2001xe} for references}
$S(M,V,Q)$ can be expanded in a similar way, and the expansion yields
\begin{equation}
S = 2^{9/4} 5^{-5/4}\pi^{1/4} \kappa^{1/2} V^{-1/4} M^{5/4}
- 2^{-11/4}5^{3/4} \pi^{5/4} \kappa^{-3/2} V^{7/4} M^{-3/4} Q^2 +O\left(Q^{4}\right)
\label{chargedSUGRA}
\end{equation}

The first terms of (\ref{chargedFT},\ref{chargedSUGRA}) are simply the zero-charge cases that have
already been discussed. Comparing the second terms we see that although the numerical coefficients
are different, at least the sign and order (i.e. having no first order term in $Q$) are
correct\footnote{Having no first order term is a consequence
of a symmetry of the entropy under $Q\rightarrow -Q$. In supergravity this symmetry is trivial;
in our field theory description this is due to the symmetry under flipping $(\mu, N,\bar{N})
\rightarrow (-\mu,\bar{N},N)$.  Since $N$ and $\bar{N}$ do not appear in the maximized entropy
formula $S(M,V,\mu)$, it is symmetric under $\mu\rightarrow -\mu$, and so after replacing $\mu$ by
$Q$ we get the $Q\rightarrow -Q$ symmetry.}.

\end{document}